\documentclass[10pt]{article}
\usepackage{amsmath}
\usepackage{amssymb}
\usepackage{graphicx}
\usepackage{cite}
\usepackage{color}
\usepackage[labelfont=bf,labelsep=period,justification=raggedright]{caption}

\makeatletter
\renewcommand{\@biblabel}[1]{\quad#1.}
\makeatother
\newcommand{\be}{\begin{equation}}
\newcommand{\ee}{\end{equation}}
\newcommand{\bea}{\begin{eqnarray}}
\newcommand{\eea}{\end{eqnarray}}
\newcommand{\rme}{{\rm{e}}}

\newcommand{\iGA}{{\iota}}
\newcommand{\BoxOp}{\partial}

\newcommand{\citep}{\cite}
\newcommand{\citet}{\cite}

\input{tcilatex}

\begin{document}

\date{}

% Title must be 150 characters or less

\begin{flushleft}
{\Large \textbf{Revisiting special relativity: A natural algebraic alternative to Minkowski spacetime} } 
% Insert Author names, affiliations and corresponding author email.
\newline
James M.~Chappell$^{\text{*}}$, Azhar Iqbal, Nicolangelo Iannella, Derek Abbott 
\newline
School of Electrical and Electronic Engineering, University of
Adelaide, South Australia 5005, Australia \newline
$\ast$ E-mail: james.m.chappell@adelaide.edu.au
\end{flushleft}

\section*{Abstract}

Minkowski famously introduced the concept of a space-time continuum in 1908, merging the three dimensions of space with an imaginary time dimension $ i c t $, with the unit imaginary producing the correct spacetime distance $ x^2 - c^2 t^2 $, and the results of Einstein's then recently developed theory of special relativity, thus providing an explanation for Einstein's theory in terms of the structure of space and time.  As an alternative to a planar Minkowski space-time of two space dimensions and one time dimension, we replace the unit imaginary $ i = \sqrt{-1} $, with the Clifford bivector 
$ \iota = e_1 e_2 $ for the plane that also squares to minus one, but which can be included without the addition of an extra dimension, as it is an integral part of the real Cartesian plane with the orthonormal basis $ e_1 $ and $ e_2 $. 
We find that with this model of planar spacetime, using a two-dimensional Clifford multivector, the spacetime metric and the Lorentz transformations follow immediately as properties of the algebra. This also leads to momentum and energy being represented as components of a multivector and we give a new efficient derivation of Compton's scattering formula, and a simple formulation of Dirac's and Maxwell's equations.  Based on the mathematical structure of the multivector, we produce a semi-classical model of massive particles, which can then be viewed as the origin of the Minkowski spacetime structure and thus a deeper explanation for relativistic effects. We also find a new perspective on the nature of time, which is now given a precise mathematical definition as the bivector of the plane.

\section*{Introduction} \label{intro}	

It has been well established experimentally that the Lorentz transformations, provide a correct translation of space and time measurements from one inertial frame of reference to another.  They were developed by Lorentz \citet{Lorentz1904} with further refinements by Poincar{\'e} \citet{poincara1905dynamique,poincare1906dynamique}, to explain the null result of the Michelson-Morley experiment, proposing a length contraction of a laboratory frame of reference moving with respect to a hypothetical aether \citep{ErnstHsu2001,goldberg1967henri,logunov2004henri,leveugle2004relativité,Voigt1887}.
Einstein, however, rederived the Lorentz transformations on the basis of two new fundamental postulates \citet{Einstein1905}, of the invariance of the laws of physics and the invariance of the speed of light, between inertial observers, thus eliminating the need for an aether. 
Minkowski in 1908, however, also derived the Lorentz transformations from a different perspective, postulating a spacetime continuum, from which the results of special relativity also naturally followed \citep{SexlUrbantke2001}, but which additionally provided a general structure for spacetime within which all the laws of physics should be described \citep{Walter1999,Zeeman1964}.
Specifically, he introduced a four-dimensional Euclidean space with the expected Pythagorean distance measure $ s^2 =  x_1^2 + x_2^2 + x_3^2 + x_4^2 $, defining $ x_4 = i c t $, where $ i = \sqrt{-1} $ is the unit imaginary, which thus allowed one to view spacetime as a conventional Euclidean space with no difference in treatment between the $ x,y,z $ and $  i c t $ coordinates \citep{minkowski1909raum,rowe2009look}, but still recovering the invariant distance measure $ s^2 = x_1^2 + x_2^2 + x_3^2 - c^2 t^2 $.  This idea was received favorably by Einstein, and by the wider scientific community at the time \citep{einstein1921relativity}, but more recently, with the desire to remain consistent with the real metric of general relativity, the unit imaginary has been replaced with a four-dimensional metric signature $ (+,+,+,-) $ \citep{TaylorWheeler,MisnerThorneWheeler1973}.

In this paper however we propose an alternate spacetime framework to Minkowski, using the multivector of a two-dimensional Clifford algebra, replacing the unit imaginary representing an imaginary time coordinate, with the Clifford bivector $ \iota = e_1 e_2 $ of the plane, defined by the orthonormal elements $ e_1 $ and $ e_2 $, which also has the property of squaring to minus one.  The bivector however has several advantages over the unit imaginary in that (i) it is a composite algebraic component of the plane and so an extra Euclidean-type dimension is not required and (ii) the bivector is an algebraic element embedded in a strictly real space, and hence consistent with the real space of general relativity.  
Clifford's geometric algebra of two-dimensions can be adopted as a suitable algebraic framework to describe special relativity, because the Lorentz transforms act separately on the parallel and perpendicular components of vectors relative to a boost direction thereby defining a two-dimensional space. 

Clifford algebra has been used previously to describe spacetime \citep{Hestenes111,Hestenes2003,Pavsic2003,zeni1992thoughtful}, however these approaches follow Minkowski in describing a four-dimensional spacetime framework with an associated mixed metric, such as the STA of Hestenes \citet{Hestenes111} which uses the four algebraic non-commuting basis elements $ \gamma_0 \dots \gamma_3 $, with $ \gamma_0^2 = -1 $ representing the time dimension and $ \gamma_1^2 = \gamma_2^2 = \gamma_3^2 =1 $ for space.  In order to relate these definitions to our framework, we can make the identifications $ e_1 = \gamma_1 \gamma_0, e_2 = \gamma_2 \gamma_0, \iota = e_1 e_2 = \gamma_1 \gamma_2 $. However the STA framework in two dimensions requires three unit vectors, as opposed to two in our approach, as well as the requirement for a mixed metric. A related approach by Baylis \citep{baylis2004relativity}, called the Algebra of Physical Space (APS), in two dimensions involves just two space unit vectors which are added to a scalar variable representing time, that is $ t + x e_1 + y e_2 $.  This is an effective approach, though we now need to define a special form for the dot product in order to return the invariant distance, whereas in our approach we achieve this from the intrinsic properties of the algebra and a definition of a spacetime event in Eq.~(\ref{spacetimeEvent}).

The representation of time with a Cartesian-type dimension in conventional approaches including STA, appears ill founded physically though due to the observed non-Cartesian like behavior of time, such as the time axis possessing a negative signature \footnote{Recall that although time is usually described by a positive Cartesian axis, it has a negative contribution to the Pythagorean distance in this space.} and the observed inability to freely move within the time dimension as is possible with space dimensions.  Our approach on the other hand requires a minimal two dimensional Euclidean space, without the need for an imposed mixed metric structure, as the invariant spacetime interval arises naturally from the properties of the algebra, with the four-vectors and tensors typically employed in special relativity replaced with the multivector, thus requiring only a single Lorentz transformation operator, which also allows Lorentz covariance to be more easily ascertained.  Also, with time now modeled as a bivector we find an algebraic structure that more appropriately models the nature of time.

Clifford's geometric algebra was first published in 1873, extending the work of Grassman and Hamilton, creating a single unified real mathematical framework over Cartesian space, which naturally included the algebraic properties of scalars, complex numbers, quaternions and vectors into a single entity, called the multivector \citep{Doran2003}.  We find that this general algebraic entity, as part of a real two-dimensional Clifford algebra $ {\rm{Cl}}_{2,0}(\Re) $, provides a natural alternative to a planar Minkowski vector space $ \Re^{2,1} $ \citep{rodrigues2007many,matolcsi1984models}.

\subsection*{Two-dimensional Clifford algebra}

In order to describe a planar space, Clifford defined two algebraic elements $ e_1 $ and $ e_2 $, with the product rule
\be \label{orthonormality}
e_1^2 = e_2^2  =  1 ,
\ee
with the composite element $ \iota = e_1 e_2 $, denoted by the Greek letter iota, being anticommuting, that is $ e_1 e_2 = - e_2 e_1 $, and assuming associativity squares to minus one \citep{Doran2003}, that is, $ \iota^2 = (e_1 e_2)^2 = e_1 e_2 e_1 e_2 = - e_1 e_1 e_2 e_2 = -1 $, and hence can be used as an alternative to the scalar imaginary $ i = \sqrt{-1} $ as a representation for the square root of minus one. A general Clifford multivector can be written through combining the various algebraic elements, as
\be \label{generalMultivector}
a + x_1 e_1 + x_2 e_2  + \iota b ,
\ee
where $ a $ and $ b $ are real scalars, $ \mathbf{x} = x_1 e_1 + x_2 e_2 $ represents a planar vector, with $ x_1, x_2 $ real scalars, and $ \iota $ is the bivector, defining an associative non-commuting algebra.  Denoting $ \bigwedge \Re^{2}  $ as the exterior algebra of $ \Re^2 $ which produces the space of multivectors $ \Re \oplus \Re^2 \oplus \bigwedge^2 \Re^2 $, a four-dimensional real vector space denoted by $ {\rm{Cl}}_{2,0} $.

\subsubsection*{Geometric product}

A key property of Clifford's algebra, is given by the product of two vectors, which are special cases of multivectors defined in Eq.~(\ref{generalMultivector}). Given the vectors $ \mathbf{u} = u_1 e_1 + u_2 e_2 $ and $ \mathbf{v} = v_1 e_1 + v_2 e_2  $, then using the distributive law for multiplication over addition, as assumed for an algebraic field, we find
\be \label{VectorProductExpand2DInitial}
\textbf{u} \textbf{v} = (u_1 e_1 + u_2 e_2)( v_1 e_1 + v_2 e_2) = u_1 v_1 + u_2 v_2  + (u_1 v_2 - u_2 v_1 ) e_1 e_2 ,
\ee
using the properties defined in Eq.~(\ref{orthonormality}).  We identify $ u_1 v_1 + u_2 v_2 $ as the dot product and $ (u_1 v_2 - u_2 v_1 ) e_1 e_2 $ as the wedge product, giving
\be \label{VectorProductExpand2D}
\textbf{u} \textbf{v} = \textbf{u} \cdot \textbf{v}  + \textbf{u} \wedge \textbf{v} .
\ee
Hence the algebraic product of two vectors produces a union of the dot and wedge products, with the significant advantage that this product now has an inverse operation.  For $ \hat{\mathbf{u}} $ and $  \hat{\mathbf{v}} $ unit vectors, we have $ \hat{\textbf{u}} \cdot \hat{\textbf{v}} = \cos \theta $ and $ \hat{\textbf{u}} \wedge \hat{\textbf{v}} = \iota \sin \theta $, we therefore have $ \hat{\textbf{u}} \hat{\textbf{v}} = \cos \theta + \iota \sin \theta $, where $ \theta $ is the angle between the two vectors.

We can see from Eq.~(\ref{VectorProductExpand2D}), that for the case of a vector multiplied by itself, that the wedge product will be zero and hence the square of a vector $ \mathbf{v}^2 = \mathbf{v} \cdot \mathbf{v} = v_1^2 + v_2^2  $, becomes a scalar quantity.
Hence the Pythagorean length of a vector is simply $ \sqrt{\mathbf{v}^2} $, and so we can find the inverse vector 
\be
\mathbf{v}^{-1} =  \frac{\mathbf{v}}{\mathbf{v}^2}.
\ee

We define the distance measure or metric over the space as the scalar part of the geometric product, which for the special case of two vectors reduces to the dot product as shown in Eq.~(\ref{VectorProductExpand2DInitial}). 

\subsubsection*{Rotations in space}

Euler's formula for complex numbers, carries over unchanged for the bivector $ \iota $,  with which we define a rotor
\be \label{EulerGA2D}
R = \cos \theta + \iota \sin \theta  = \rme^{ \iota \theta}  ,
\ee
which produces a rotation by $ \theta $ on the $ e_1 e_2 $ plane, in the same way as rotations on the Argand diagram.  For example, for a unit vector $ \mathbf{v} = e_1 $ along the $ e_1 $ axis, acting with the rotor from the right we find $ \mathbf{v} R  =  e_1 (\cos \theta + \iota \sin \theta) = \cos \theta e_1 + e_2 \sin \theta $, thus describing an anti-clockwise rotation by $ \theta $.  If we alternatively act from the left with the rotor, we will find a clockwise rotation by $ \theta $.

However, we now show, that a rotation can be described more generally as a sequence of two reflections. Given a vector $ \mathbf{n}_1 $ normal to a reflecting surface, with an incident ray given by $ \mathbf{I} $, then we find the reflected ray \citep{Doran2003}
\be \label{reflectedRay}
\mathbf{r} = - \mathbf{n}_1 \mathbf{I} \mathbf{n}_1.
\ee
If we apply a second reflection, with a unit normal $ \mathbf{n}_2 $, then we have
\be
\mathbf{r} = \mathbf{n}_2 \mathbf{n}_1 \mathbf{I} \mathbf{n}_1 \mathbf{n}_2 = ( \cos \theta - \iota \sin \theta ) \mathbf{I} ( \cos \theta + \iota \sin \theta ) =\rme^{-\iota \theta} \mathbf{I} \rme^{\iota \theta} ,
\ee
using Eq.~(\ref{VectorProductExpand2D}) for two unit vectors.  If the two normals  $ \mathbf{n}_1 $ and  $ \mathbf{n}_2 $ are parallel, then no rotation is produced.  In fact the rotation produced is twice the angle between the two normals.

Hence rotations are naturally produced by conjugation, where if we seek to rotate a vector $ \mathbf{v} $ by an angle $ \theta $, we calculate
\be \label{quaternionRotation}
\mathbf{v}'  = \rme^{-\iota \theta/2} \mathbf{v} \rme^{\iota \theta/2} ,
\ee
which rotates in an anticlockwise direction.  The rotation formula in Eq.~(\ref{quaternionRotation}) above, can in two-space, be simplified to a single right acting operator $ \mathbf{v}'  =  \mathbf{v} \rme^{\iota \theta} $.  However this simplification is only possible in two-dimensions for the special case of rotations on vectors, and will be incorrect when applied to other algebraic elements or to vectors in higher dimensions, and hence Eq.~(\ref{quaternionRotation}) is the preferred way to apply operators such as rotors on vectors and multivectors.

\section*{Results}

\subsection*{Clifford multivectors as a framework for space and time}

Considering Minkowski's definition of spacetime coordinates and Eq.~(\ref{generalMultivector}), we describe planar spacetime events as the multivector  
\be \label{spacetimeEvent}
X =  (x_1 e_1 +  x_2 e_2 ) \iota + \iota c t  =  \mathbf{x} \iota + \iota c t = (c t + \mathbf{x} ) \iota ,
\ee
with $  \mathbf{x} $ representing the position vector in the plane and $ t $ the observer time.  This is without loss of generality for planar collisions, as we can always orientate this plane to lie in the plane of the relative velocity vector between the frames, and special relativity only requires two axes, the orthogonal and parallel directions to the relative velocity vector. The interpretation of a coordinate in Eq.~(\ref{spacetimeEvent}) is the conventional one, of an observer moving through a preconfigured coordinate system, which at each point has a properly synchronized clock, from which the moving observer can read off the other frames local time $ t $ and position $ \mathbf{x} $ \citep{TaylorWheeler}.  We then find the spacetime interval to be
\be \label{invariantInterval}
X^2 = ( \mathbf{x} + c t  )\iota ( \mathbf{x} + c t ) \iota =   \mathbf{x}^2 - c^2 t^2 + c  t  \mathbf{x} \iota + c t \iota \mathbf{x} = \mathbf{x}^2 - c^2 t^2 ,
\ee
using the fact that $ \iota $ anticommutes with each component of $ \mathbf{x} $, and $ \iota^2 = -1 $, giving the correct spacetime distance. It is of interest to note that a modified spacetime coordinate given by  $ X = \mathbf{x} + \iota c t $ will also give the invariant spacetime distance as shown in Eq.~(\ref{invariantInterval}), however using the definition in Eq.~(\ref{spacetimeEvent}), we find that both the spacetime coordinates and the electromagnetic field have the identical Lorentz transformation, as well as enabling us to provide a unified description of the Dirac and Maxwell equations as shown in Eq.~(\ref{Dirac3DTimeOne}) and Eq.~(\ref{MaxwellPrimitiveOne}). 

We have from Eq.~(\ref{spacetimeEvent}) the multivector differential
\be \label{SpaceTimeDifferential}
d X = ( c d t + d \mathbf{x} ) \iota ,
\ee
which is independent of space and time translations as required by the principle of relativity and so can describe the larger Poincar{\'e} group.  For the rest frame of the particle we have $ d X_0^2 = - c^2  d \tau^2 $, where we define in this case $ t $ to represent the proper time $ \tau $ of the particle. We have assumed that the speed $ c $ is the same in the rest and the moving frame, as required by Einstein's second postulate.  Now, if the spacetime interval defined in Eq.~(\ref{invariantInterval}) is invariant under the Lorentz transformations defined later in Eq.~(\ref{HomLorentzGroup}), then we can equate the rest frame interval to the moving frame interval, giving
\be \label{findingGamma}
c^2 d \tau^2 = c^2 d t^2 - d \mathbf{x}^2 = c^2 d t^2 - \mathbf{v}^2 d t^2 = c^2 d t^2 \left ( 1 - \frac{\mathbf{v}^2}{c^2 } \right ) ,
\ee
with $ d \mathbf{x} = \mathbf{v} d t $, and hence, taking the square root, we find the time dilation formula $ d t = \gamma d \tau $ where
\be \label{gamma}
\gamma = \frac{1}{\sqrt{1 - \mathbf{v}^2/c^2 }} .
\ee

From Eq.~(\ref{SpaceTimeDifferential}), we can now calculate the proper velocity, differentiating with respect to the proper time, giving the velocity multivector
\be \label{velocityMultivector}
U = \frac{d X}{d \tau} =  \frac{d \mathbf{x}}{d t} \frac{d t}{d \tau} \iota + \iota c \frac{d t}{d \tau}   =  \gamma \mathbf{v} \iota + \gamma c \iota = \gamma \left ( c + \mathbf{v} \right ) \iota,
\ee
where we use $ \frac{d t}{d \tau} = \gamma $ and $ \mathbf{v} = \frac{d \mathbf{x}}{d t} $. 
We then find
\be
U^2 =  (\gamma \mathbf{v} \iota + \gamma c \iota )^2  = \left (\frac{1}{1-\mathbf{v}^2/c^2} \right ) ( \mathbf{v}^2 - c^2) = - c^2 .
\ee
We define the momentum multivector
\be \label{momentumMultivector}
P = m U = \gamma m \mathbf{v} \iota +  \gamma m c \iota =  \mathbf{p} \iota +  \frac{E}{c} \iota ,
\ee
with the relativistic momentum $ \mathbf{p} = \gamma m \mathbf{v} $ and the total energy $ E = \gamma m c^2 $.

Now, as $ U^2 = - c^2 $, then $ P^2 = - m^2 c^2 $ is an invariant between frames describing the conservation of momentum and energy, which gives
\be \label{MultivectorEnergyConservation}
P^2 c^2 = \mathbf{p}^2 c^2 - E^2 = - m^2 c^4 ,
\ee
the relativistic expression for the conservation of momentum-energy.  
The square of the velocity multivector resolving to a constant $ -c^2 $ gives the expected property for the acceleration multivector $ A = \frac{d U}{d \tau } $, of being orthogonal the the velocity multivector, from
\be
\frac{ d }{d \tau} U^2 = 2 U \cdot \frac{ d U }{ d \tau} = 2 U \cdot A = 0
\ee
using the chain rule from geometric calculus \citep{GA}.

\subsubsection*{The Lorentz Group}

The Lorentz transformations describe the transformations for observations between inertial systems in relative motion. The set of  transformations describing rotations and boosts connected with the identity are described as proper and is referred to as the restricted Lorentz group described in four-dimensional spacetime as $ \rm{SO}^+(3,1) $, whereas if we also permit reflections we expand the transformations to the homogeneous Lorentz group $ \rm{SO}(3,1) $.  Its worth noting though that in two-dimensions reflections are also part of the restricted Lorentz group.

The most general transformation of a coordinate multivector is given by
\be
X' = L X N ,
\ee
where $ L $ and $ N $ are general multivectors, with the coordinate multivector $ X $ defined in Eq.~(\ref{spacetimeEvent}).
Requiring the invariance of the spacetime distance given by $ X^2 $ we find the relation
\be \label{distanceEquation}
X'^2 = L X N L X N = X^2 ,
\ee
which is satisfied if $ N L = L N = \pm 1 $.  For a general multivector given by $ L = a + \mathbf{v} + \iota t $, if we define the dagger operation $ L^{\dagger} = a - \mathbf{v} - \iota t $, then we produce a scalar $ L L^{\dagger} = L^{\dagger} L = a^2 - \mathbf{v}^2 + t^2 $. Hence in Eq.~(\ref{distanceEquation}) we require $ N  = L^{\dagger} $ with $ L L^{\dagger} = \pm 1 $.  For the case  $ L L^{\dagger} =  + 1 $, we can write $ L = \rme^B $, where $ B = \phi \hat{\mathbf{v}} + \iota \theta $, see Appendix, which describes a set of transformations connected with the identity. Though these transformations are not closed they nevertheless satisfy $ L L^{\dagger} = \rme^B \rme^{B^{\dagger}}= \rme^B \rme^{-B} = \rme^0 = 1 $ as required, using the fact that a multivector commutes with itself and naturally describes the Thomas rotation for two non-parallel boosts, that is $ \rme^{ \phi_1 \hat{v}_1 } \rme^{ \phi_2 \hat{v}_2 } = \rme^{\phi \hat{\mathbf{v}} + \iota \theta} $. In order to close the operators consisting of general boosts and rotations we need to write $ L = \rme^{ \phi \hat{v} } \rme^{ \iota \theta } $. Other special transformations can be considered, such as with $ N = L $ provided we enforce the condition $ a = 0 $, which then describes space and time reflections, so that we can write a unit multivector $ L = \hat{B} $, where $ \hat{B} = B/|B| $ and $ |B|= \sqrt{B^2} = \sqrt{\phi^2 - \theta^2 } $ , giving $ X' = \hat{B} X \hat{B} $, see Eq~(\ref{reflectedRay}).  The second general case  $ L L^{\dagger} = -1 $ can be represented as $ \hat{B} \rme^{ \phi \hat{B} } $ which is a combination of a proper Lorentz boost and a reflection and so not part of the restricted Lorentz group, but useful in representing collision processes with an associated energy transfer such as photons reflecting off electrons as in Compton scattering, described in Eq.~(\ref{ComptonVector}).

The exponential of a multivector is defined by constructing the Taylor series
\be \label{exponentialMultivector}
\rme^M = 1 + M +\frac{M^2}{2!} + \frac{M^3}{3!} + \dots
\ee
which is absolutely convergent for all multivectors $ M $ \citep{Hestenes111}.
Also because of the closure of multivectors under addition and multiplication, we see that the exponential of a multivector, must also produce another multivector, and we find, in fact, a unique multivector $ L = \rme^{M} $, for each multivector $ M $ \citep{Hestenes111}.  
Hence in summary, all operators of the form 
\be \label{HomLorentzGroup}
L = \rme^{\phi \hat{\mathbf{v}} } \rme^{\iota \theta } ,
\ee
applied to the multivector $ M $ using the transformation
\be \label{coordinateBoosts}
M' = L M L^{\dagger} = \rme^{\phi \hat{\mathbf{v}} } \rme^{\iota \theta } M \rme^{-\iota \theta } \rme^{-\phi \hat{\mathbf{v}} } ,
\ee
will leave the spacetime distance invariant, defining the restricted Lorentz group \citep{zeni1992thoughtful}. We find for $ \phi = 0 $ pure rotations as described by Eq.~(\ref{quaternionRotation}), and for $ \theta = 0 $, we find pure boosts, where $ M $ can denote the coordinate, momentum or electromagnetic field multivectors.

\subsubsection*{Spacetime boosts}

Using the first component of the restricted Lorentz group defined in Eq.~(\ref{HomLorentzGroup}), operators of the form $ \rme^{\phi \hat{\mathbf{v}} } $, where the vector $ \mathbf{v} = v_1 e_1 + v_2 e_2 \mapsto \phi \hat{\mathbf{v}} $, where $ \hat{\mathbf{v}} $ is a unit vector, with $ \hat{\mathbf{v}}^2 =1  $, we find
\be \label{exponentiateVector}
 \rme^{ \phi \hat{\mathbf{v}} } = 1 + \phi  \hat{\mathbf{v}} + \frac{\phi^2}{2!} + \frac{\phi^3 \hat{\mathbf{v}} }{3!} + \frac{\phi^4}{4!} + \dots = \cosh \phi + \hat{\mathbf{v}} \sinh \phi .
\ee
Transforming the spacetime coordinates $ X =  \mathbf{x} \iota + \iota c t  $ we find
\bea \label{XBoostCoshSinh}
X' & = & \rme^{-\hat{\mathbf{v}} \phi/2  } ( \mathbf{x} \iota  + \iota c t  ) \rme^{ \hat{\mathbf{v}} \phi/2 } \\ \nonumber
& = &  \rme^{ - \hat{\mathbf{v}} \phi} \mathbf{x}_{ll} \iota + \mathbf{x}_{\perp} \iota +  \iota c t \rme^{ \hat{\mathbf{v}} \phi }  \\ \nonumber
& = &  (\cosh \phi |x_{ll}| - c t \sinh \phi ) \hat{\mathbf{v}} \iota + x_{\perp} \iota + \iota  (c t \cosh \phi  - \sinh \phi |x_{ll}| )  , \nonumber
\eea
where $ x_{ll} $ and $ x_{\perp} $ are the coordinates parallel and perpendicular respectively to the boost velocity direction $ \hat{\mathbf{v}} $, which is the conventional Lorentz boost, in terms of the rapidity $ \phi $, defined by $ \tanh \phi = v/c $, which can be rearranged to give  $ \cosh \phi = \gamma $ and $ \sinh \phi = \gamma v /c $.  Substituting these relations we find
\be \label{XBoost}
X'  = \gamma ( |x_{ll}| - v t ) \hat{\mathbf{v}} \iota + \mathbf{x}_{\perp} \iota  + \iota  \gamma \left ( c t  - \frac{ v |x_{ll}|}{c} \right ) ,
\ee
which thus gives the transformation $  x_{ll}' = \gamma (|x_{ll}| - v t ) $, $ x_{\perp}' = x_{\perp} $ and $ c t' = \gamma (c t - \frac{v |x_{ll}|}{c} ) $, the correct Lorentz boost of coordinates.  
The formula in Eq.~(\ref{coordinateBoosts}) can be simply inverted to give $ X = \rme^{ \hat{\mathbf{v}} \phi /2 } X' \rme^{ -\hat{\mathbf{v}} \phi /2 } $, using the fact that $ \rme^{ \hat{\mathbf{v}} \phi /2 } \rme^{- \hat{\mathbf{v}} \phi /2 } = \rme^0 = 1 $.
The relativity of simultaneity is a fundamental result of special relativity, and from the perspective of the Clifford multivector Eq.~(\ref{spacetimeEvent}), we see that it stems from the fact that during a boost operation, the terms for space $ e_1 $ and $ e_2 $ become mixed, resulting in the bivector term $ e_1 e_2 $, thus creating a variation in the observers time coordinate.  Similarly the momentum multivector, shown in Eq.~(\ref{momentumMultivector}), will follow the same transformation law between frames shown in Eq.~(\ref{coordinateBoosts}), with $ P' = L P L^{\dagger} $. Serendipitously, we also find that the Lorentz boost of electromagnetic fields is subject to the same operator as coordinate transformations given by Eq.~(\ref{coordinateBoosts}). 

Given a general electromagnetic field represented by the multivector $ F = E_x e_1 + E_y e_2 + \iota c B = \mathbf{E} + \iota c B $, where for two-dimensional space we only have available a single magnetic field direction $ B_z $ out of the plane, represented by the axial vector $ \iota B = i e_3 B_z $.  Applying the boost according to Eq.~(\ref{coordinateBoosts}), with the exponentiation of a general boost vector $ \mathbf{v} \mapsto  \phi \hat{\mathbf{v}} $, we find
\bea \label{exponentiateVectorField}
\rme^{ - \frac{\hat{\mathbf{v}} \phi}{2} } F \rme^{ \frac{ \hat{\mathbf{v}} \phi}{2} } &  = & \left (\cosh \frac{\phi}{2} - \hat{\mathbf{v}} \sinh \frac{\phi}{2} \right ) \left (\mathbf{E}_{ll} + \mathbf{E}_{\perp} + \iota c B \right ) \left (\cosh \frac{\phi}{2} + \hat{\mathbf{v}} \sinh \frac{\phi}{2} \right ) \\ \nonumber
& = & \mathbf{E}_{ll} + \mathbf{E}_{\perp} \left (\cosh \phi + \hat{\mathbf{v}} \sinh \phi \right ) + \iota c B  \left (\cosh \phi + \hat{\mathbf{v}} \sinh \phi \right ) \\ \nonumber
& = & \mathbf{E}_{ll} + \gamma \left ( \mathbf{E}_{\perp} + \iota \mathbf{v} B \right )  + e_1 e_2  \left ( \gamma c B - \frac{|\mathbf{E}_{\perp}| \gamma v}{c} \right ) ,  \nonumber
\eea
which are the correct Lorentz transformations for an electromagnetic field.
That is, the parallel field $ \mathbf{E}_{ll} $ is unaffected, the perpendicular field $ \mathbf{E}_{\perp} $ has been increased to $ \gamma \mathbf{E}_{\perp} $ and the term $ e_1 e_2 |\mathbf{E}_{\perp}| \gamma v/c  $, represents the $ e_1 e_2 $ plane, also describable with an orthogonal vector $ e_3 $ in three-space, hence this term gives the expected induced magnetic field $ B_z $ from the perpendicular electric field $ \mathbf{E}_{\perp} $.  

Hence the exponential map of a Clifford multivector, naturally produces the restricted Lorentz transformations of spacetime coordinates and the electromagnetic field in the plane using the Lorentz boost Eq.~(\ref{coordinateBoosts}), with the spacetime coordinate multivector given by Eq.~(\ref{spacetimeEvent}) and the field multivector $ F = \mathbf{E} + \iota c B $.   

\subsubsection*{Velocity addition rule}

If we apply two consecutive parallel boosts, $ \mathbf{v}_1 = v_1 \hat{\mathbf{v}} \mapsto \phi_1 \hat{\mathbf{v}} $ and $ \mathbf{v}_2 = v_2 \hat{\mathbf{v}} \mapsto \phi_2 \hat{\mathbf{v}} $, where $ \tanh \phi = \frac{v}{c} $, we have the combined boost operation
\be
\rme^{ \phi_1 \hat{\mathbf{v}} } \rme^{ \phi_2 \hat{\mathbf{v}} } = \rme^{ (\phi_1+\phi_2) \hat{\mathbf{v}} }.
\ee
Hence we have a combined boost velocity
\be \label{velocityAddition}
v = c \tanh (\phi_1+\phi_2 ) = \frac{\tanh \phi_1 + \tanh \phi_2 }{ 1 + \tanh \phi_1 \tanh \phi_2 } = \frac{v_1 + v_2}{1+ v_1 v_2/c^2 },
\ee
the standard relativistic velocity addition formula. By inspection, the velocity addition formula implies that a velocity can never be boosted past the speed $ c $, which confirms $ c $ as a speed limit.

Hence, we have now demonstrated from the ansatz of the spacetime coordinate described by the multivector shown in Eq.~(\ref{spacetimeEvent}), that we produce the correct Lorentz transformations, where the variable $ c $ is indeed found to be an invariant speed limit.
Numerically therefore, $ c $ can be identified as the speed of light, since this is the only known physical object which travels at a fixed speed and represents a universal speed limit.  

\subsection*{Applications} \label{Apps} 

\subsubsection*{$ \pi^{+} $-meson decay}

A classic example of experimental confirmation for the special theory of relativity is its application to the decay of $ \pi^{+} $-mesons, which are observed to enter the atmosphere at high velocity $ \mathbf{v} $ from outer space, having a known decay time at rest of $ \tau_{\pi} = 2.55 \times 10^{-8} $ s, giving a spacetime coordinate multivector at rest of $ X = \iota c \tau_{\pi} $. Boosting these coordinates to the $ \pi^{+} $-meson velocity, we have a boost $ \rme^{ \iota \hat{\mathbf{v}} \phi /2 } $, where $ \tanh \phi = v/c $, so we therefore find from Eq.~(\ref{coordinateBoosts})
\be
X' = R X R^{\dagger} = \rme^{ - \hat{\mathbf{v}} \phi /2 } \iota c \tau_{\pi} \rme^{ \hat{\mathbf{v}} \phi /2 } = \iota c \tau_{\pi} \rme^{ \hat{\mathbf{v}} \phi } = \iota c \tau_{\pi} ( \cosh \phi + \hat{\mathbf{v}} \sinh \phi ) = \gamma \mathbf{v} \tau_{\pi} \iota  + \iota \gamma c \tau_{\pi} .
\ee
So that we have a decay time in laboratory coordinates of $ c t =  \gamma c \tau_{\pi} $, with a track length in the laboratory of $ \mathbf{x} = \gamma \mathbf{v} \tau_{\pi} $, in agreement with experimental determinations \citep{French1987}.

\subsubsection*{Doppler shift}

The Doppler shift of light, refers to the change of frequency caused by the relative velocity between source and observer.
In the rest frame of the source, we can describe a single wavelength $ \lambda $ of emitted light using Eq.~(\ref{spacetimeEvent}), setting up the $ e_1 $ axis along the line of sight, as
\be
X = \left ( c T + \lambda e_1 \right ) \iota  = \left ( \lambda + \lambda e_1 \right ) \iota ,
\ee
where $ T = \lambda/c $ is the period of the wave, which gives $ X^2 = 0 $ as required for a photon. We can describe an observer in relative motion with a boost in the $ \hat{\mathbf{v}} = e_1 $ direction using $ \tanh \phi = v/c $, and we find from Eq.~(\ref{coordinateBoosts})
\be
X' =  \rme^{ - \hat{\mathbf{v}} \phi /2 } \left ( \lambda e_1 \iota + \lambda \iota \right )  \rme^{ \hat{\mathbf{v}} \phi /2 } =  \left ( \lambda e_1 \iota + \lambda \iota \right )  \rme^{ \hat{\mathbf{v}} \phi } = \gamma \lambda \left ( 1- \frac{v}{c} \right ) e_1 \iota + \gamma \lambda \left ( 1 - \frac{v}{c} \right ) \iota .
\ee
So using the space(or alternatively time) component we find $ \lambda' = \lambda \gamma  \left ( 1- \frac{v}{c} \right ) $ and using $ c = f \lambda $ we find the standard relativistic Doppler shift formula 
\be
\frac{f'}{f} = \frac{1}{\gamma \left (1 - \frac{v}{c} \right ) } = \frac{\sqrt{1+v/c}}{\sqrt{1-v/c}}.
\ee

\subsubsection*{Thomas rotation}

A surprising result occurs when we apply two non-parallel boosts, followed by their inverse boosts, in that the velocity of the frame does not return to zero. Furthermore, there is a rotation of the frame, called the Thomas rotation, a result, in fact, not noticed until 1925 \citep{TaylorWheeler}.

For the case of two consecutive general boosts given by
\be \label{ThomasRotation}
R = \rme^{ - \phi_2 \hat{\mathbf{v}}_2/2}  \rme^{ - \phi_1 \hat{\mathbf{v}}_1/2 } = \rme^{ -  \phi_c \hat{\mathbf{v}}_c/2 } \rme^{  - \iota \theta/2 } ,
\ee
where we use the results from Appendix A, to write this in terms of a single combined boost $ \phi_c \hat{\mathbf{v}}_c $ and a rotation $ \theta $, finding, using the results from Appendix A,
\be
\tan \frac{\theta}{2} = \frac{ \sin \delta \sinh \frac{\phi_1}{2} \sinh \frac{\phi_2}{2} }{\cos \delta \sinh \frac{\phi_1}{2} \sinh \frac{\phi_2}{2} - \cosh \frac{\phi_1}{2} \cosh \frac{\phi_2}{2} } ,
\ee 
where $ \delta $ is the angle between the boost directions, given by $ \cos \delta = \hat{\mathbf{v}}_1 \cdot \hat{\mathbf{v}}_2 $. Hence we can see that only for parallel boosts, that is $ \delta = 0 $, will there not in fact be a Thomas rotation $ \theta $, of the frame.

We can also write the Thomas rotation as a single exponential of a multivector
\be
R  = \rme^{ - \phi_t \hat{\mathbf{v}}_t/2 - \iota \theta_t/2 },
\ee
using the results of Appendix B.

\subsubsection*{Scattering processes}

It is well established that energy and momentum conservation applies in relativistic dynamics, provided that the rest energy $ m c^2 $ is now included along with the appropriate relativistic corrections, that is, defining momentum as $ \gamma m \mathbf{v} $, and the energy as $ \gamma m c^2 $. We now show that the two conservation laws can be bundled into a single momentum multivector defined in Eq.~(\ref{momentumMultivector}), giving a new perspective on momentum and energy conservation as the conservation of a multivector.

For example, if we are given a set of particles that are involved in an interaction, which then produce another set of particles as output. Then, in order to describe this collision interaction process we firstly include a separate momentum multivector for each particle, and then energy and momentum conservation between the initial and final states is defined by
\be
\sum P_{{\rm{initial}}} = \sum P_{{\rm{final}}} ,
\ee
assuming we are dealing with an isolated system.
We know $ E = | \mathbf{p} | c $ for a massless particle, so using Eq.~(\ref{momentumMultivector}) we write the momentum multivector for a photon as $ \Gamma = \mathbf{p} \iota + \iota | \mathbf{p} | $, which gives $ \Gamma^2 = 0 $ and for a massive particle $ P^2 = - m^2 c^2 $ as shown in Eq.~(\ref{MultivectorEnergyConservation}).

\begin{figure}[htb]

\begin{center}
\includegraphics[width=3.4in]{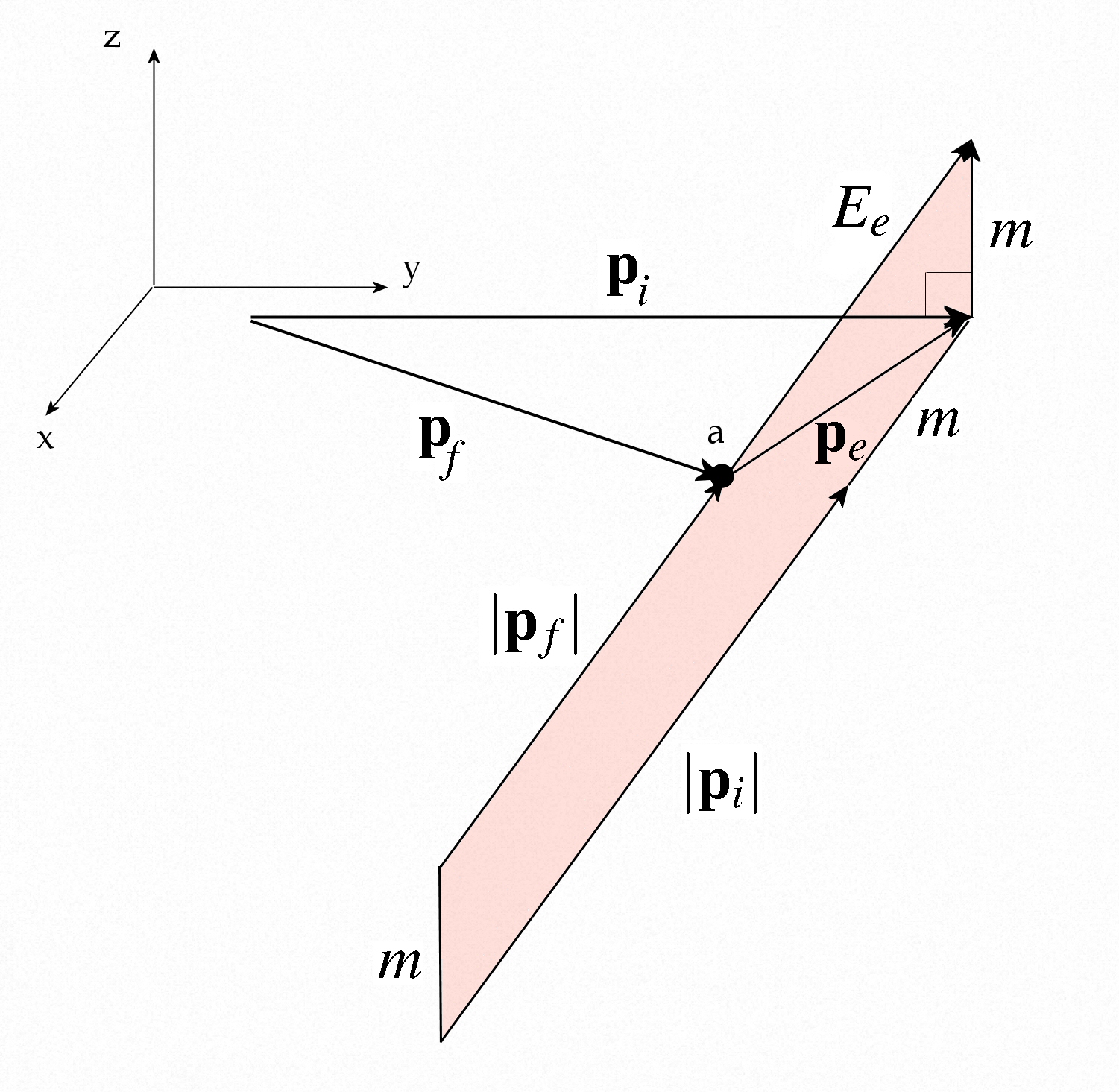}
\end{center}

\caption{Graphical solution to Compton scattering (natural units with $ c = 1 $).  In order to obtain possible experimental outcomes the point $a$ is moved in the plane of $x$ and $y$, as shown, which automatically satisfies conservation of momentum given by the vector triangle, $ \mathbf{p}_i  = \mathbf{p}_f + \mathbf{p}_e $ and the locus of points which also maintains the shape of the figure in the vertical plane as a parallelogram (shown in red) satisfies the conservation of energy. We have the Pythagorean distance giving the final energy of the electron $ E_e = \sqrt{\mathbf{p}_e^2 + m^2 } $, so that the requirement of a parallelogram implies the conservation of energy $ |\mathbf{p}_i| + m = |\mathbf{p}_f| + E_e $. Hence this 3D graphical solution simultaneously satisfies the relativistic conservation of momentum and energy providing the solutions for Compton scattering.  \label{ComptonGraphical}}
\end{figure}

For Compton scattering, which involves an input photon striking an electron at rest, with the deflected photon and moving electron as products, we can write energy and momentum conservation using the multivectors as $ \Gamma_i + P_i = \Gamma_f + P_f $, which we can rearrange to 
\be \label{ComptonMomEnergy}
(\Gamma_i -\Gamma_f ) + P_i = P_f .
\ee
Squaring both sides we find
\be \label{SquaringComptonMomEnergy}
(\Gamma_i -\Gamma_f )^2 + P_i(\Gamma_i -\Gamma_f ) + (\Gamma_i -\Gamma_f ) P_i + P_i^2 = P_f^2 ,
\ee
remembering that in general the multivectors do not commute. Now, we have the generic results that $ P_i^2 = P_f^2 = - m^2 c^2 $ and $ (\Gamma_i -\Gamma_f )^2 = \Gamma_i^2 + \Gamma_f^2 - \Gamma_i \Gamma_f - \Gamma_f \Gamma_i = - 2 \Gamma_i \cdot \Gamma_f = - 2 (\mathbf{p}_i \cdot  \mathbf{p}_f - | \mathbf{p}_i | | \mathbf{p}_f |) = 2 | \mathbf{p}_i | | \mathbf{p}_f | (1 - \cos \theta ) $, using $ \Gamma_i^2 = \Gamma_f^2 = 0 $. For the following two terms in Eq.~(\ref{SquaringComptonMomEnergy}), using $ P_i = \iota m c $, we have $ m c ( \iota (\Gamma_i -\Gamma_f ) + (\Gamma_i -\Gamma_f )\iota ) = - 2 m c (| \mathbf{p}_i |  - | \mathbf{p}_f | ) $.  We therefore find from Eq.~(\ref{SquaringComptonMomEnergy}) that
\be
 | \mathbf{p}_i | | \mathbf{p}_f | (1 - \cos \theta ) - m c (| \mathbf{p}_i|  - | \mathbf{p}_f| ) = 0 .
\ee
Dividing through by $ | \mathbf{p}_i | | \mathbf{p}_f | $ and substituting $ |\mathbf{p}| = \frac{h}{\lambda} $ we find Compton's well known formula
\be \label{ComptonStandard}
\lambda_f - \lambda_i = \frac{h}{ m c} (1 - \cos \theta ) . 
\ee
The advantage of the momentum multivector is that energy and momentum
conservation can be considered in unison as shown in Eq.~(\ref{ComptonMomEnergy}), which also provides a clear solution path, whereas typical textbook methods rely on manipulating two separate equations describing momentum and energy conservation \citep{French1987}. The multivector equation shown in Eq.~(\ref{ComptonMomEnergy}) also leads to a graphical solution, shown in Fig.~\ref{ComptonGraphical}.   This 3D visual model allows us to find a solution while simultaneously conserving relativistic momentum and energy.

We can also describe this process using GA as firstly the reflection of the photon off the electron, given by $ P' = -\hat{\mathbf{v}}  P  \, \hat{\mathbf{v}} $ using Eq.~(\ref{reflectedRay}), followed by a deboost of the photon due to the energy lost to the electron, given by the operator $ \rme^{-\phi \hat{\mathbf{v}}/2} $, so that the new photon momentum multivector will be given by 
\be \label{ComptonVector}
P' = - \rme^{-\phi \hat{\mathbf{v}}/2} \hat{\mathbf{v}} P \, \hat{\mathbf{v}} \rme^{\phi \hat{\mathbf{v}}/2} ,
\ee
where $ \hat{\mathbf{v}} = \cos \delta e_1 + \sin \delta e_2 $ is the unit vector defining the direction of the electrons recoil with $ \delta $ measured from the same axis as $ \theta $, and $ \phi $ represents the amount of deboost of the photon, given by $ \gamma = \cosh \phi = \frac{\kappa- \cos \delta \sqrt{ \kappa^2 - \sin^2 \delta}}{\sin^2 \delta} $, where $ \kappa = \frac{| \mathbf{p}_f |}{| \mathbf{p}_i |} $ using $ | \mathbf{p}_f | $ calculated from Eq.~(\ref{ComptonStandard}) using the relation $ \cot \frac{\theta}{2} =  \left (1+ \frac{| \mathbf{p}_i |}{m_e c} \right )\tan \delta  $ .  While $ \phi $ needs to be calculated using the analysis leading to Eq.~(\ref{ComptonStandard}), Eq.~(\ref{ComptonVector}) nevertheless gives us an intuitive and coordinate free way to describe the photon in the Compton effect, as a reflection and deboost.

\subsection*{Modeling fundamental particles as multivectors}

In the previous section we found that the momentum multivector provides a natural description for Compton scattering involving the interaction of photons and electrons, and so guided by the mathematical structure of the multivector we produce a simple model for the electron producing results consistent with special relativity. 
Using the multivector defined in Eq.~(\ref{waveMultivector}), we can represent a particle moving with a velocity $ \mathbf{v} $ as
\be \label{electronMultivector}
P =  \hbar \mathbf{k} \iota  + \iota \frac{\gamma \hbar \omega_0}{2 c} ,
\ee
where $ \hbar \mathbf{k} =  \gamma m \mathbf{v} $. For a particle at rest, we therefore have $ P_0  = \iota \frac{\hbar \omega_0}{2 c} = \iota \frac{E}{c} $, where we use the de~Broglie relation between total energy and frequency $ E = \hbar \omega $, to find $ \omega_0 = \frac{ 2 m c^2}{\hbar} $.
The bivector $ \iota $ can be interpreted as a rotation operator, and so for a simplified semi-classical-type model, we can assume a circular periodic motion with a radius
\be \label{Electron_r0}
r_0 = \frac{\hbar}{2 m_e c } = \frac{\lambda_c}{2},
\ee
where $ \lambda_c = \frac{\hbar}{m_e c } $ is the reduced Compton wavelength, which then gives the tangential velocity $ v = r_0 \omega_0 =  \left (\frac{\hbar}{2 m_e c } \right ) \left ( \frac{ 2 m_e c^2}{\hbar} \right ) = c $ indicating an orbiting lightlike particle.
This model leads to a natural explanation for time dilation, using the proper time invariant distance $ -c^2 d \tau^2 = d \mathbf{x}^2 - c^2 d t^2 $, which can be rearranged to $ c^2 d t^2 = d \mathbf{x}^2 + c^2 d \tau^2 $, then because the proper time distance given by the circumference always moves perpendicular to the momentum vector, due to the bivector $ \iota $ being perpendicular to the plane, then the net path distance of the lightlike particle, representing the observed time $ c d t $ is simply the Pythagorean distance $ d \mathbf{x}^2 + c^2 d \tau^2 $ and because all photons are measured with the same speed according to special relativity, the period of the orbit will be increased by $ \gamma $ giving the expected time dilation effect.

We have now arrived at a model similar to previous elementary models of the electron developed by various authors \citep{hestenes1990zitterbewegung,vaz1993zitterbewegung,pavsic1993spin}.  The models are based on the {\it{zitterbewegung}} phenomena, first described by \citet{Schrodinger1930}, an effect recently verified by experiment   \citep{wunderlich2010quantum,gerritsma2010quantum,zawadzki2011zitterbewegung}.  
Schr{\"o}dinger interpreted the {\it{zitterbewegung}} as arising from the interference of positive and negative energy states, but later described by \citet{Penrose2004} as a lightlike particle oscillating at the speed of light, with an amplitude equal to the reduced Compton wavelength. 

In the footsteps of previous investigations \citep{Schrodinger1930, Penrose2004, hestenes1990zitterbewegung,pavsic1993spin,rodrigues1993zitterbewegung,vaz1993zitterbewegung,vaz1995barut,vaz1995spinning}, a future development is to extend this work to three dimensional space.

\subsection*{Wave mechanics}

A further application of the momentum multivector defined in Eq.~(\ref{momentumMultivector}), is through the standard substitutions $ \mathbf{p} = - \iota \hbar \nabla $ and $ \mathbf{E} = \iota \hbar \partial_t $, from which we produce the spacetime gradient operator as
\be \label{BoxGradient}
\BoxOp = \iota \left ( \partial_t + \nabla   \right ) ,
\ee
where $ \nabla = e_1 \partial_x + e_2 \partial_y  $ is the two-space gradient operator.  We then find $ -\BoxOp^2 = \partial_t^2 - \nabla^2 $ the d'Alembertian in two dimensions, so that $ \BoxOp $ is a square root of the d'Alembertian. We therefore write for the Dirac equation
\be \label{Dirac3DTimeOne}
\BoxOp \psi  = m  \psi ,
\ee
where $ \psi $ is a general multivector, shown in Eq.~(\ref{generalMultivector}), which gives a Lorentz covariant equation isomorphic to the conventional Dirac equation in two dimensions (see Appendix), and comparable to the Dirac equation previously developed in three dimensional Clifford algebra \citep{hestenes2003mysteries,Boudet}.  
Acting from the left a second time with the differential operator $ \BoxOp  $ upon Eq.~(\ref{Dirac3DTimeOne}) we produce the Klein-Gordon equation, $ \left (\partial_t^2 - \nabla^2  \right ) \psi = -m^2 \psi $ as required. 

Taking Eq.~(\ref{Dirac3DTimeOne}) with $ m = 0 $ and adding a source multivector $ J = ( \rho + \mathbf{J} ) \iota $, we can write
\be \label{MaxwellPrimitiveOne}
\BoxOp \psi = J,
\ee
which is isomorphic to Maxwell's equations in two dimensions, provided we write the electromagnetic field as the multivector $ \psi = \mathbf{E} + \iota B $ \citep{Boudet}. The square of the field produces the Lorentz invariant $ \psi^2 = \mathbf{E}^2 - B^2 $.  If we seek to complete the current multivector $ J $ to a full multivector with a bivector term $ \iota s $, that is $ J = ( \rho + \mathbf{J} + \iota s ) \iota $, then we find that $ \iota s $ represents magnetic monopole sources.
It is straightforward to show Lorentz covariance. Beginning with the primed frame we have from Eq.~(\ref{MaxwellPrimitiveOne}) $ \partial' \psi' = J' $.  However we have $ \psi' = L \psi L^{\dagger} $ and $ J' = L J L^{\dagger} $, which implies 
\be
L \partial L^{\dagger} L \psi L^{\dagger} = L J L^{\dagger} ,
\ee
which implies therefore implies $ \partial F = J $, thus demonstrating covariance, using the property of the Lorentz transformation that
$ L L^{\dagger} = L^{\dagger} L = 1 $.

If we calculate
\be \label{FJEqn}
\psi J = \left ( \mathbf{E} + \iota B \right ) \left ( \rho + \mathbf{J} \right ) \iota = \left ( \mathbf{J} \cdot \mathbf{E} + \rho \mathbf{E} + \iota  \mathbf{J} B  + \iota \rho B + \mathbf{E} \wedge \mathbf{J} \right ) \iota = \left ( W + \mathbf{f} + \iota \rho B + \mathbf{E} \wedge \mathbf{J}  \right ) \iota ,
\ee
then, inside the bracket, we find the work done by the field on the current $ W = \mathbf{J} \cdot \mathbf{E} $ as a scalar and the vector force on the charges as $ \mathbf{f} =  \rho \mathbf{E} + \iota  \mathbf{J} B $, equivalent to $ \mathbf{f} =  \rho \mathbf{E} + \mathbf{J} \times \mathbf{B} $ in three dimensions.
We can write this in terms of the field alone through substituting Eq.~(\ref{MaxwellPrimitiveOne}), which gives
\be \label{FJFieldsEqn}
\iota \psi \left ( \partial \psi \right ) =  \partial_t u + \nabla \cdot \mathbf{S} - \partial_t \mathbf{S} - \nabla u + ( \mathbf{E} \cdot \nabla + \nabla \cdot \mathbf{E} ) \mathbf{E} +  \left ( \mathbf{E} \cdot \nabla  - \nabla \cdot \mathbf{E} \right ) \iota B 
\ee
where we have written $ u = \frac{1}{2} \left ( \mathbf{E}^2 + B^2 \right ) $ representing the field energy and $ \mathbf{S} = \iota B \mathbf{E} $ the Poynting vector in two dimensions.  Inspecting expressions Eq.~(\ref{FJEqn}) and Eq.~(\ref{FJFieldsEqn}) we can see that it expresses the conservation of energy and momentum.  In fact it is convenient to define a field momentum multivector
\be \label{stressEnergyTensor}
T = \frac{1}{2} \psi \iota \psi^{\dagger} = \left ( \frac{1}{2} \left ( \mathbf{E}^2 + B^2 \right ) + \iota B \mathbf{E} \right ) \iota = \left ( u + \mathbf{S} \right ) \iota  ,
\ee
which is in the form of a momentum multivector, as defined in Eq.~(\ref{momentumMultivector}).
Now, we see that the first four terms in Eq.~(\ref{FJFieldsEqn}) can be expressed as $ \partial T = \iota \left ( \partial_t + \nabla   \right ) \left ( u + \mathbf{S} \right ) \iota = - \partial_t u  - \nabla \cdot \mathbf{S} + \partial_t \mathbf{S} + \nabla u - \nabla \wedge \mathbf{S} $, therefore we can express the conservation of energy as $ \partial \cdot T = -\iota \psi \cdot J $ which gives  $ \partial_t u + \nabla \cdot \mathbf{S} = -\mathbf{J} \cdot \mathbf{E} $, or Poynting's theorem for the conservation of energy. The conservation of charge $ \partial_t \rho + \nabla \cdot \mathbf{J} = 0 $ also follows from Maxwell's equation through taking the divergence of Eq.~(\ref{MaxwellPrimitiveOne}).

An simple solution path is found through defining the field $ \psi $ in terms of a multivector potential $ A = \iota \left ( -V + c \mathbf{A} + \iota M \right ) $, with $ M $ describing a possible monopole potential, given by $ \psi = \BoxOp A $.
We then find Maxwell's equations defined in Eq.~(\ref{MaxwellPrimitiveOne}) in terms of a potential becomes $ \BoxOp \left ( \BoxOp A \right )  = \BoxOp^2 A = J $ and because $ \BoxOp^2 = \nabla^2 - \frac{1}{c^2} \partial_t^2  $ is a scalar differential operator we have succeeded in separating Maxwell's equations into four independent inhomogeneous wave equations, given by the scalar, vector and bivector components of the multivectors, each with known solution.  

For the Dirac equation, using the definition of Eq.~(\ref{stressEnergyTensor}) to define the Dirac current, we find defining a general Dirac wave function as $ \psi =  \lambda + \mathbf{E} + \iota B $, then
\be
T  =  \frac{1}{2} \left (\lambda + \mathbf{E} + \iota B \right )(\lambda + \mathbf{E} - \iota B) \iota  =  \frac{1}{2} \left ( \lambda^2 + \mathbf{E}^2 + B^2 \right ) \iota + ( \lambda \mathbf{E} + \iota B \mathbf{E} ) \iota =  \left ( u + \mathbf{S} \right ) \iota , 
\ee
then we find a positive definite density $ u $ and a vector $ \mathbf{S} $.  Then we find the divergence gives a conserved current  $ \BoxOp \cdot J = \iota \left ( \partial_t + \nabla   \right ) \cdot   \left ( u + \mathbf{S} \right ) \iota  = - \left ( \partial_t  u + \nabla  \cdot \mathbf{S}  \right ) = 0 $ as required, now appearing as the conservation of energy.

It is known that Einstein's equations for general relativity describing gravity, if placed within a (2+1) spacetime, does not allow the propagation of gravitational waves as they require two orthogonal degrees of freedom orthogonal to the direction of propagation.  Although, it should be noted, that Witten showed that the equations of GR can still describe the global topology of a (2+1) spacetime.

\section*{Discussion} \label{conclusion} 

It is well established that Clifford's geometric algebra, is a natural formalism suited for the study of geometrical operations of the plane, such as reflections and rotations \citep{Doran2003}.  However, we demonstrate additionally that spacetime represented as the Clifford multivector, as shown in Eq.~(\ref{spacetimeEvent}), is a natural alternative to Minkowski spacetime, producing the correct spacetime interval and the required Lorentz transformation, directly from the properties of the algebra.  Also the use of the momentum multivector defined in Eq.~(\ref{momentumMultivector}) allows the principle of momentum and energy conservation to be interpreted as the conservation of a multivector. We also find that the momentum multivector leads to a unified description of the Dirac and Maxwell's equations in the plane.  The mathematical structure of the wave multivector in Eq.~(\ref{waveMultivector}), also leads to a simple model for the internal structure of the electron, in accordance with previous developments \citep{Penrose2004,hestenes1990zitterbewegung,pavsic1993spin,rodrigues1993zitterbewegung,vaz1993zitterbewegung,vaz1995barut,vaz1995spinning}.  

The definition of a spacetime event as a multivector in Eq.~(\ref{spacetimeEvent}), also provides a new perspective on the nature of time, in that rather than being defined as an extra Euclidean-type dimension, it becomes instead a composite quantity of space, the bivector $ e_1 e_2 $.  Minkowski's famous quote is therefore particularly apt, \emph{Henceforth space by itself, and time by itself, are doomed to fade away into mere shadows, and only a kind of union of the two will preserve an independent reality} \citep{Minkowski}. 
As we have seen in Eq.~(\ref{EulerGA2D}), a bivector represents a rotation operator, and so it is natural to interpret time as an angular rotation at the de~Broglie frequency $w = \frac{E}{\hbar} $ related to the frequency of the  {\it{zitterbewegung}} \citep{Schrodinger1930,Penrose2004,hestenes1990zitterbewegung}, and encapsulated by a two dimensional model implied from the multivector description in Eq.~(\ref{electronMultivector}), shown in Fig.~{\ref{ElectronBivector}}. 

A view of time as a rotational entity, has also been supported by recent experiments, which have identified a fluctuating electric field at the de~Broglie frequency for an electron \citep{catillon2008search}, and the use of the rotating electric field in circularly polarized light as an attosecond clock to probe atomic processes \citep{Pfeiffer2011,Ueda2011,eckle2008attosecond,RevModPhys81163}. Hence the popular notion of time, as the `river of time',  certainly based in part on the pronouncement of Newton in the Principia, Book 1 \citep{Newton1686}, that time \emph{\dots flows equably without relation to anything external \dots }, combined with time being promoted by Minkowski as a fourth dimension, may perhaps need to be amended to a description of a rotational entity, and adopting a water analogy, time would therefore be viewed descriptively as a whirlpool or an eddy current. Newton's concept of the steady \emph{flow of time} would relate in the multivector model to the constant spin rate at the de~Broglie frequency of each particle that is constant in the particles' rest frame, thus indeed \emph{flowing equably}. Unforeseen by Newton though was the observed variation in this rotation rate with an \emph{external} observer in relative motion, which produces the relativistic effects identified by Einstein.

It is also now interesting to consider the impact on the nature of time if we expand the two-dimensional multivector in Eq.~(\ref{spacetimeEvent}) to three dimensions.
This firstly allows space vectors to possess three degrees of freedom $ (x,y,z) $, but also the single bivector $ \iota = e_1 e_2 $ representing time will now expand to include the three bivectors of a three-dimensional multivector.  Hence a direct implication of representing time as a bivector is that when expanding this model to three dimensions, time will now become three dimensional \citep{tifft1996three,lehto1990periodic,Cole1980comments,weinberg1980some}, but in our case associated with the three rotational degrees of freedom of three-dimensional space.

Minkowski spacetime diagrams, consisting of a space axis and a time axis are still applicable, though the time axis no longer represents a Euclidean time dimension, but simply shows the algebraic relationship between time as a bivector and space as a vector.  The abstract nature of Minkowski diagrams are indeed confirmed by the rotation of the coordinate axes for a moving observer, which are tilted with respect to the original frame when displayed on the Minkowski diagram, a practice which is purely formal and not indicating a real rotation of the space or time axes between the frames \citep{French1987}.  Boosts are conventionally interpreted as rotations in time in comparisons to rotations in space.  However this interpretation needs to be revised from the new perspective of Clifford multivectors, with spatial rotations seen as bivector operators of the form $ \rme^{\iota \theta } $ and boosts as vector operators of the form $ \rme^{\phi \hat{\mathbf{v}} } $.

There are many definitions of clock time possible, such as the rotation of the earth on its axis, or the vibration of a quartz crystal, however the one discussed here, based on the bivector rotation of particles is perhaps the most fundamental.  The arrow of time is another property of time, however it has been recognized previously that this arises from the universe being far from equilibrium in a low entropy thermodynamic state.  The steady progress towards high entropy as required by the second law of thermodynamics leading to the `heat death' of the universe gives a perceived direction to time, though this is essentially unrelated to the definition of time given by the bivector rotation.  The definition of time, as a bivector representing rotation, also allows the difficult concept of time beginning with the big bang to be more accessible as it now simply implies the non existence of rotational degrees of freedom before the big bang.  The creation of time with the big bang is in agreement with many philosophical conceptions of time, such as Augustine's statement, \emph{The world was made, not in time, but simultaneously with time} \citep{Augustine1972}.

In summary, this approach from an abstract mathematical perspective based on the ansatz of spacetime represented by a Clifford multivector shown in Eq.~(\ref{spacetimeEvent}), produces the correct spacetime metric and Lorentz transformations directly from the properties of the algebra, and thus similar to Minkowski's approach, we explain the two postulates of Einstein based on the geometrical structure  of spacetime.  This systematic approach, is also shown to be advantageous in describing the Lorentz transformations, in that an exploration of the exponential map of a multivector, naturally produced  rotations, boosts and the Thomas rotation of frames, and in fact the restricted Lorentz group represented simply as the multivector exponentials $ \rme^{ \phi \hat{\mathbf{v}}} \rme^{\iota \theta} $.  This Lorentz transform operator is generic, as it simultaneously provides the transformation for the coordinate, momentum-energy and electromagnetic fields, with all these objects modeled uniformly as multivectors. This can be compared with the conventional approach that uses four-vectors to represent coordinates and the momentum-energy but with a different structure, the antisymmetric field tensor, used to represent the electromagnetic fields, with necessarily different transformation operations for each type of object. 
Hence we see significant pedagogical benefits with the use of multivectors as a description of spacetime, which allow the Lorentz transformations as well as the Dirac equation and Maxwell's equations, to arise naturally in a simplified algebraic setting, without any unnecessary mathematical `overheads', such as matrices, four-vector, complex numbers, tensors or metric structures. It is hoped with the simplified two dimensional framework using only real numbers and two algebraic entities $ e_1, e_2 $, that a greater fundamental understanding of quantum mechanical processes at a fundamental level may be possible.
The minimalist system that we have presented having just sufficient complexity to describe special relativity is therefore in line with Einstein's ideal that: \emph{It can scarcely be denied that the supreme goal of all theory is to make the irreducible basic elements as simple and as few as possible without having to surrender the adequate representation of a single datum of experience} \citep{einstein1934method}.

%\section{Acknowledgements} \label{ack} We would like to thank... 	

\newpage

\section*{Analysis}

\subsection*{Boost-rotation form of a multivector}

Given a general two-space multivector written as
\be \label{MultivectorM}
M = r \cos \alpha + s \cos \beta e_1 + s \sin \beta e_2 + \iota r \sin \alpha  ,
\ee
then requiring $ M M^{\dagger} = 1 $, we find the condition $ r^2 - s^2 = 1 $.
If we seek to write Eq.~(\ref{MultivectorM}) in the exponential form 
\bea \label{exponentialForm}
\rho \rme^{ \phi \hat{\mathbf{v}} } \rme^{ \theta \iota } & = & \rho \Big ( \cosh \phi \cos \theta + \sinh \phi (v_1 \cos \theta - v_2 \sin \theta ) e_1 \\ \nonumber
& & + \sinh \phi (v_2 \cos \theta + v_1 \sin \theta ) e_2 + \iota \cosh \phi \sin \theta  \Big ),  \nonumber
\eea
consisting of a separate boost and rotation, we require
\bea \label{exponentialRelations}
\rho & = & \pm \sqrt{r^2 -s^2 } = \pm 1, \,\, \theta = \alpha , \,\, \phi = {\rm{arctanh}} \left (\frac{s}{r} \right ) =  {\rm{arctanh}} \left (\frac{s}{\sqrt{1+s^2}} \right ) \\ \nonumber
v_1 & = & \cos ( \beta - \alpha)  \, , \,\,\, v_2 = \sin ( \beta - \alpha ) ,  \nonumber
\eea
which are all well defined.

\subsection*{Exponential of a multivector}

It is found that exponentiating the even subalgebra $ a + \iota b $, which is a closed subalgebra, produces rotations and dilations, while exponentiating the vector $ \mathbf{v} = v_1 e_1 + v_2 e_2 $, produces Lorentz boosts.  However the odd subalgebra is not closed and consequently the set of boosts is not closed but can also involve the Thomas rotation. However if we use the exponential of a full multivector, we encompass the set of non-parallel boosts and Thomas rotations. 

Firstly, defining $ B = \mathbf{v} + \iota b $, then $ B^2 = (\mathbf{v} + \iota b)^2 = \mathbf{v}^2 - b^2 + b \mathbf{v} \iota + b \iota \mathbf{v} =  \mathbf{v}^2 - b^2 $, a scalar.  We also have the result that $ \rme^{a+B} = \rme^a \rme^B $, as $ a $ is a scalar and hence commutes with $ B $.  Hence
\bea \label{derivationExpMultivector}
\rme^{a + \mathbf{v} + \iota b} & = & \rme^{a} \rme^{\mathbf{v} + \iota b} \\ \nonumber
& = & \rme^{a} \left ( 1 + B + \frac{\mathbf{v}^2 - b^2 }{2!}+ \frac{B (\mathbf{v}^2 - b^2  )}{3!} + \frac{(\mathbf{v}^2 - b^2)^2 }{4!} + \dots  \right ) \\ \nonumber
& = & \rme^{a} \Big ( 1 +  \frac{\mathbf{v}^2 - b^2  }{2!} +  \frac{(\mathbf{v}^2 - b^2  )^2}{4!} + \dots \\ \nonumber
& & + \frac{B}{\sqrt{\mathbf{v}^2 - b^2 }} \Big ( \sqrt{\mathbf{v}^2 - b^2 } + \frac{\sqrt{\mathbf{v}^2 - b^2 } (\mathbf{v}^2 - b^2  )}{3!} + \dots \Big )  \Big ) \\ \nonumber
& = & \rme^{a}  \left ( \cosh | B | + \hat{B} \sinh | B |   \right ) , \nonumber
\eea
where $  | B | = |\sqrt{\mathbf{v}^2 - b^2}| $, assuming $ \mathbf{v}^2>b^2 $, and $ \hat{B} = \frac{B}{|B|} = \frac{\mathbf{v} + \iota b}{|B|} $.
If $ v^2 < b^2 $ we simply replace the hyperbolic trigonometric functions with trigonometric functions, and if $ \mathbf{v}^2 = b^2 $, then referring to the second line of the above derivation, we see that all terms following $ B $ are zero, and so, in this case $ \rme^{a + \mathbf{v} + \iota b} = \rme^{a} (1+ \mathbf{v} + \iota b ) $.  Hence the exponential of a general multivector $ \rme^M = \rme^{a + \mathbf{v} + \iota b} $ is well defined for all $ a $, $ \mathbf{v} $ and $ b $.
The reverse process, of finding the exponent for a given multivector, the logarithm of a multivector, is not always defined.

We are now in a position to classify the various transformation operators as shown in Table \ref{tableFormalisms}.  

\begin{table}
	\centering

\begin{tabular}{|l|l|l|l|}
\hline
Operation $ L $  & Description & Comments   \\
\hline \hline
$ \rme^{\iota \theta/2 } $ & Exp. of bivector, rotn  by $ \theta $  & Group of rotations  \\
$ \rme^{\phi \hat{\mathbf{v}}/2 } \rme^{ \iota \theta/2 } $ & Compound boost rotn. &  Lorentz group connected with identity  \\
$ a + \mathbf{v} + \iota \theta $  & General multivector & General boost, rotation, reflection, dilation  \\
\hline
$ \rme^{\phi \hat{\mathbf{v}}/2 } $ & Boost with $ \tanh \phi = v/c $ &  Pure boosts  \\
$ \rme^{\phi \hat{\mathbf{v}}/2 + \iota \theta/2 } $ & Exp. of multivector  &  Multiple boosts (Thomas rotn.)  \\
\hline
\end{tabular}
\caption{Classification of the Lorentz group, with the Lorentz transformation defined by $ X' = L X L^{\dagger} $, with $ L L^{\dagger} = \pm 1 $. The first section forms a chain of subgroups, with each succeeding group having the previous groups as subgroups. The entries in the second section do not form a group, but are useful in describing pure boosts and compound boost situations. }
\label{tableFormalisms}
\end{table}

\subsection*{Geometric calculus}

The product rule for differentiation using Clifford variables 
\be
\nabla (A B) = e_i \partial_i (A B ) = e_i (\partial_i A) B + e_i  A (\partial_i B ),
\ee
where we respect non-commutivity of the algebraic parts $ e_i $ of the differential operator.

The chain rule for a general function on a multivector can be expressed
\be
\partial_x M(f) = \partial_f M \cdot \partial_x f,
\ee
where the dot product is specified on the RHS.
For example for $ M = \left (x e_1 + x^3 e_2 \right )^2 $ then defining $ \mathbf{f} = x e_1 + x^3 e_2 $, we have $ M = \mathbf{f}^2 $ and so we find
\be
\partial_x M = 2 \mathbf{f} \cdot \partial_x \mathbf{f} = 2 (x e_1 + x^3 e_2 ) \cdot (e_1 + 3 x^2 e_2 ) = 2 x + 6 x^5 .
\ee
This can be checked by expanding $ M(\mathbf{f}) = \mathbf{f}^2 = x^2 + x^6 $ and then $ \partial_x M = 2 x + 6 x^5 $ as required.

We can generalise the chain rule for the gradient operator
\be
\nabla M(f) = \dot{\nabla} \left ( \partial_f M \cdot \dot{f} \right ) ,
\ee
for example, for $ M = \mathbf{E}^2 $, we have
\be
\nabla \left ( \mathbf{E}^2 \right ) =\dot{\nabla} \left ( 2 \mathbf{E} \cdot  \dot{\mathbf{E}} \right ) = 2 \left ( \nabla \cdot \mathbf{E}  - \mathbf{E} \wedge \nabla \right ) \mathbf{E}.
\ee
This can be confirmed by expanding the following expressions
\be
\nabla \left ( \mathbf{E}^2 \right ) = e_1 \left ( 2 E_x \partial_x E_x + 2 E_y \partial_x E_y \right ) + e_2 \left ( 2 E_x \partial_y E_x + 2 E_y \partial_y E_y \right )
\ee
\be
2 \left ( \nabla \cdot \mathbf{E} \right ) \mathbf{E} = e_1 \left ( 2 E_x \partial_x E_x + 2 E_x \partial_y E_y \right ) + e_2 \left ( 2 E_y \partial_x E_x + 2 E_y \partial_y E_y \right )
\ee
\be
2 \left ( \mathbf{E} \wedge \nabla \right ) \mathbf{E} = e_1 \left ( 2 E_x \partial_y E_y - 2 E_y \partial_x E_y \right ) + e_2 \left ( 2 E_y \partial_x E_x - 2 E_x \partial_y E_x \right ).
\ee
We also have for the gradient operator
\be
\nabla \cdot (f \mathbf{E} ) = ( \nabla f ) \cdot \mathbf{E} + f \nabla \cdot \mathbf{E}
\ee
\be
\nabla (f \mathbf{E} ) = ( \nabla f ) \mathbf{E} + f \nabla \mathbf{E} .
\ee
The following expressions are also useful
\be
\mathbf{w} \cdot ( \iota \mathbf{E} ) = ( \mathbf{w} \iota ) \cdot \mathbf{E}
\ee
\be
\mathbf{E} \wedge ( \mathbf{w} \iota ) = ( \iota \mathbf{E} ) \wedge \mathbf{w} = \iota ( \mathbf{E} \cdot \mathbf{w} ) .
\ee

\subsubsection*{The Dirac equation in two dimensions}

We now aim to describe a form of the Dirac equation in Clifford algebra isomorphic to the conventional Dirac equation
\be
\gamma^{\mu} \partial_{\mu} | \psi \rangle + i q \gamma^{\mu} A_{\mu} | \psi \rangle = - i m | \psi \rangle ,
\ee
where $ i = \sqrt{-1} $ and $ \gamma^{\mu} $ are the Dirac matrices, using natural units in which $ c = \hbar = 1 $.
If we reduce the number of spatial dimensions to two, then the Dirac algebra can fit within the Pauli algebra, and we can write the Dirac equation as
\be \label{Dirac2DBook}
i \partial_t | \psi \rangle =  - i \left ( \sigma_1 \partial_x + \sigma_2 \partial_y \right ) | \psi \rangle + \sigma_3 m | \psi \rangle,
\ee
where $ \sigma_1 , \sigma_2 , \sigma_3 $ are the Pauli matrices \citep{thaller1992dirac}.  Naturally the one-dimensional Dirac equation can be found by ignoring the $ y $ direction as $ i \partial_t | \psi \rangle =  - i \sigma_1 \partial_x | \psi \rangle + \sigma_3 m | \psi \rangle $.

If we select a spinor mapping to the two dimensional multivector as
\be \label{SpinorMapping2DMultivector}
| \psi \rangle = \begin{bmatrix} a_0 + j a_3  \\ a_2 + j a_1  \end{bmatrix} \leftrightarrow \psi = a_0 + a_2 e_1 + a_1 e_2 + a_3 e_1 e_2 ,
\ee
then we find the following mapping  for the Pauli matrices
\be
\sigma_k | \psi \rangle  \leftrightarrow  e_k \psi  
\ee
for $ k = 1,2 $ and
\be
i \sigma_3 | \psi \rangle = \sigma_1 \sigma_2 | \psi \rangle \leftrightarrow  e_1 e_2 \psi  
\ee
using $ i I = \sigma_1 \sigma_2 \sigma_3 $.
Expanding Eq.~(\ref{Dirac2DBook}) we find
\be
- \partial_t \psi = \partial_x \sigma_1 \psi + \partial_y \sigma_2 \psi + m \sigma_1 \sigma_2 \psi
\ee
using the relation $ i \sigma_3 = \sigma_1 \sigma_2 $.  Mapping this to the multivector defined in Eq.~(\ref{SpinorMapping2DMultivector}) we find
\be
- \partial_t \psi = e_1 \partial_x \psi + e_2 \partial_y \psi + m  e_1 e_2 \psi.
\ee
Multiplying from the left by $ -e_1 e_2 $ we find
\be
e_1 e_2 \left ( \partial_t + \nabla \right ) \psi =  m \psi.
\ee
Hence defining $ \BoxOp = \iGA \left ( \partial_t + \nabla \right ) $, where $ \iGA = e_1 e_2 $, we find
\be \label{DiracGAAppendix}
\BoxOp \psi = m \psi
\ee
where the Dirac wavefunction is described by the multivector in Eq.~(\ref{SpinorMapping2DMultivector}).  We then find $ -\BoxOp^2 = \partial_t^2 - \nabla^2 $ the d'Alembertian, thus allowing us to recover the Klein-Gordon equation in two dimensions from Eq.~(\ref{DiracGAAppendix}).  The one dimensional Dirac equation also given by Eq.~(\ref{DiracGAAppendix}) but with the spatial gradient operator $ \nabla = e_1 \partial_x $.

\subsubsection*{Bilinear observables}

Given $ \psi =  \lambda + \mathbf{E} + \iota B $, then
\be
\psi \tilde{\psi}  = \left (\lambda + \mathbf{E} + \iota B \right )(\lambda + \mathbf{E} - \iota B)   = \lambda^2 + \mathbf{E}^2 + B^2  + 2 ( \lambda \mathbf{E} + \iota B \mathbf{E} ) =  \rho + \mathbf{v}  , 
\ee
where $ \tilde{\psi} =  \lambda + \mathbf{E} - \iota B $ is the reversion operation.  We therefore define the probability current as 
\be \label{Jdefn}
J = \psi \tilde{\psi} \iota =  \psi \iota  \psi^{\dagger} = \left (  \rho + \mathbf{v} \right ) \iota .
\ee
This is in the form of a four-velocity, from inspection of the velocity multivector in Eq.~(\ref{velocityMultivector}). The term $ \rho = \lambda^2 + \mathbf{E}^2 + B^2 $ is a positive definite scalar equivalent to $ \rho = | \psi |^2 = \langle \psi | \psi \rangle $ conventionally calculated for the probability density. Thus this equation relates the Dirac current $ J $ to the wavefunction $ \psi $. For $ \psi \rightarrow \psi S $, we find
\be 
J = \psi S \iota S^{\dagger} \psi^{\dagger}
\ee
so that provided $ S \iota S^{\dagger} = \iota $, $ S $ represents a gauge transformation.  Multiplying from the right by $ S $ and remembering that $ S^{\dagger} S $ is a scalar, then we find $ S \iota = \iota  S $ which implies that $ S $ commutes with $ \iota $ and hence $ S = \rme^{\iota \theta } = \cos \theta + \iota \sin \theta $ describing a rotation in the spin plane $ \iota = e_1 e_2 $.

Calculating the divergence of our probability current
\be \label{divcurrent}
\BoxOp \cdot J = \iota \left ( \partial_t + \nabla   \right ) \cdot   \left ( \rho + \mathbf{v} \right ) \iota  = - \partial_t  \rho - \nabla  \cdot \mathbf{v} = 0
\ee
which is recognizable expression for conservation of charge or probability.
Now with this definition of current we find
\bea \label{divv}
\nabla \cdot \mathbf{v} & = & \nabla \cdot 2 ( \lambda \mathbf{E} + \iota B \mathbf{E} ) \\ \nonumber
& = & 2 \left ( \nabla \lambda \right ) \cdot  \mathbf{E} + 2 \lambda \left ( \nabla \cdot  \mathbf{E} \right ) + 2 \left ( \nabla B \right ) \cdot \left ( \iota  \mathbf{E} \right ) + 2 B \left ( \nabla \cdot  \iota \mathbf{E} \right ) \\ \nonumber
& = & 2 \left ( \nabla \lambda \right ) \cdot  \mathbf{E} + 2 \lambda \left ( \nabla \cdot  \mathbf{E} \right ) - 2  \mathbf{E}  \cdot \left ( \iota \nabla B \right )  - 2 \iota B \left ( \nabla \wedge \mathbf{E} \right ) ,  \nonumber
\eea
where we have used $ \mathbf{v} \cdot \left ( \mathbf{w} \iota \right ) = \left ( \mathbf{v} \wedge \mathbf{w} \right ) \iota $ and $ \mathbf{v} \cdot \left ( \mathbf{w} \iota \right ) = \left ( \iota \mathbf{v} \right ) \cdot \mathbf{w} $.

To confirm our definition of probability current, we firstly write the Dirac equation and its reverse 
\bea
\partial_t \psi & = & - \nabla \psi - m \iota \psi \\ \nonumber
 \partial_t \tilde{\psi} & = & - \tilde{\psi} \nabla + m \tilde{\psi} \iota . \nonumber
\eea
Multiplying the first equation on the left with $ \tilde{\psi} $ and the second equation on the right with $ \psi $ we obtain
\bea
 \tilde{\psi} \left ( \partial_t \psi \right ) & = & -  \tilde{\psi} \left ( \nabla \psi \right ) - m  \tilde{\psi} \iota \psi \\ \nonumber
 \left ( \partial_t \tilde{\psi} \right ) \psi & = & - \left ( \tilde{\psi} \nabla \right ) \psi + m \tilde{\psi} \iota \psi . \nonumber
\eea
Adding these two equations we find
\be
 \partial_t \left ( \tilde{\psi} \psi \right ) + \tilde{\psi} \left ( \nabla \psi \right ) + \left ( \tilde{\psi} \nabla \right ) \psi = 0 .
\ee
Investigating the second and third terms, we find
\be
\tilde{\psi} \left ( \nabla \psi \right ) = (\lambda + \mathbf{E} - \iota B) \nabla (\lambda + \mathbf{E} + \iota B) = (\lambda + \mathbf{E} - \iota B) \left ( \nabla \lambda + \nabla \cdot \mathbf{E} + \nabla \wedge \mathbf{E} - \iota \nabla B \right )
\ee
\be
 \left ( \tilde{\psi} \nabla \right ) \psi = \left ( (\lambda + \mathbf{E} - \iota B) \nabla \right ) (\lambda + \mathbf{E} + \iota B) = \left ( \nabla \lambda + \nabla \cdot \mathbf{E} - \nabla \wedge \mathbf{E} - \iota \nabla B \right ) (\lambda + \mathbf{E} + \iota B)
\ee
The bivector terms $ \lambda \nabla \wedge \mathbf{E} +  \mathbf{E} \wedge \nabla \lambda -  \mathbf{E} \wedge \left ( \iota \nabla B \right ) - \iota B \nabla \cdot \mathbf{E} $ cancel, leaving the scalar and vector components. We have the scalar parts
\be
2 \left ( \nabla \lambda \right ) \cdot \mathbf{E} + 2 \lambda \left ( \nabla \cdot \mathbf{E} \right ) - 2 \mathbf{E} \cdot \left( \iota \nabla B \right ) - 2 \iota B \left ( \nabla \wedge \mathbf{E} \right )
\ee
which confirms our definition of probability current in Eq.~(\ref{Jdefn}) and Eq.~(\ref{divcurrent}), through comparison with Eq.~(\ref{divv}).

We have the vector terms
\bea
& & \partial_t \mathbf{v} + 2 \lambda \nabla \lambda + 2 \mathbf{E} \left ( \nabla \cdot \mathbf{E} \right ) - 2 B \nabla B - 2 \lambda \iota \nabla B - 2 \iota B \nabla \lambda \\ \nonumber
& = & \partial_t \mathbf{v} + \nabla \rho + 2 \left ( \mathbf{E} \wedge \nabla \right ) \mathbf{E} - 2 \nabla B^2 - 2 \iota \nabla \left ( \lambda B \right ) . \nonumber
\eea

Alternatively, using the Dirac equation and its reverse and multiplying the first equation on the right with $ \tilde{\psi} $ and the second equation on the left with $ \psi $ we obtain
\bea
 \left ( \partial_t \psi \right )  \tilde{\psi} & = & -   \left ( \nabla \psi \right ) \tilde{\psi} - m  \iota \psi \tilde{\psi} \\ \nonumber
\psi  \left ( \partial_t \tilde{\psi} \right ) & = & - \psi \left ( \tilde{\psi} \nabla \right )  + m \psi \tilde{\psi} \iota  . \nonumber
\eea
Adding these two equations we find
\be
 \partial_t \left ( \psi \tilde{\psi}  \right ) +  \left ( \nabla \psi \right ) \tilde{\psi} + \psi \left ( \tilde{\psi} \nabla \right )  + 2 \iota m \mathbf{v} = 0 .
\ee
Investigating the second and third terms, we find
\be
 \left ( \nabla \psi \right ) \tilde{\psi} = \nabla (\lambda + \mathbf{E} + \iota B) (\lambda + \mathbf{E} - \iota B)  =  \left ( \nabla \lambda + \nabla \cdot \mathbf{E} + \nabla \wedge \mathbf{E} - \iota \nabla B \right ) (\lambda + \mathbf{E} - \iota B)
\ee
\be
 \psi \left ( \tilde{\psi} \nabla \right )  = (\lambda + \mathbf{E} + \iota B)  \left ( (\lambda + \mathbf{E} - \iota B) \nabla \right ) = (\lambda + \mathbf{E} + \iota B) \left ( \nabla \lambda + \nabla \cdot \mathbf{E} - \nabla \wedge \mathbf{E} - \iota \nabla B \right ) .
\ee
The bivector terms $ \lambda \nabla \wedge \mathbf{E} +  \mathbf{E} \wedge \nabla \lambda -  \mathbf{E} \wedge \left ( \iota \nabla B \right ) - \iota B \nabla \cdot \mathbf{E} $ cancel, leaving the scalar and vector components. We have the scalar parts
\be
2 \left ( \nabla \lambda \right ) \cdot \mathbf{E} + 2 \lambda \left ( \nabla \cdot \mathbf{E} \right ) - 2 \mathbf{E} \cdot \left( \iota \nabla B \right ) - 2 \iota B \left ( \nabla \wedge \mathbf{E} \right )
\ee
which confirms our definition of probability current in Eq.~(\ref{Jdefn}) and Eq.~(\ref{divcurrent}), through comparison with Eq.~(\ref{divv}).

We have the vector terms
\bea
& & \partial_t \mathbf{v} + 2 \lambda \nabla \lambda + 2 \mathbf{E} \left ( \nabla \cdot \mathbf{E} \right ) + 2 B \nabla B + 2 \lambda \iota \nabla B - 2 \iota B \nabla \lambda + 2 \iota m \mathbf{v} \\ \nonumber
& = & \partial_t \mathbf{v} + \nabla \rho  + 2 \left ( \mathbf{E} \wedge \nabla \right ) \mathbf{E} + 2 \lambda \iota \nabla B - 2 \iota B \nabla \lambda + 2 \iota m \left (  2 ( \lambda \mathbf{E} + \iota B \mathbf{E} ) \right )  \nonumber
\eea
using the result that $ \frac{1}{2} \nabla \mathbf{E}^2 =  \left ( \nabla \cdot \mathbf{E} \right )\mathbf{E} - \left ( \mathbf{E} \wedge \nabla \right ) \mathbf{E} $.

We have the Poynting vector $ \mathbf{s} = \iota B \mathbf{E} $ which is equivalent so $ S = \mathbf{E} \times \mathbf{B} $ in three dimensions.  We have the energy density $ u = \frac{1}{2} \left ( \mathbf{E}^2 + B^2 \right ) $, so that
\bea \label{FJFieldsEqnDetailed}
\iota \psi \left ( \partial \psi \right ) & = & \iota \left ( \mathbf{E} + \iota B \right ) \iota ( \partial_t + \nabla ) \left ( \mathbf{E} + \iota B \right ) \\ \nonumber
& = & ( \mathbf{E} - \iota B ) ( \partial_t + \nabla ) \left ( \mathbf{E} + \iota B \right ) \\ \nonumber
& = & \mathbf{E} \partial_t \mathbf{E} + B \partial_t B  - \iota B (\nabla \wedge \mathbf{E} )+ (\iota \mathbf{E}) \cdot \nabla B  - \iota \partial_t B \mathbf{E}  - \iota B \partial_t \mathbf{E} - B \nabla B + \left ( \nabla \cdot \mathbf{E} \right ) \mathbf{E} \\ \nonumber
& & + \mathbf{E} \nabla \wedge \mathbf{E} - \iota B \nabla \cdot \mathbf{E} - \mathbf{E} \wedge ( \iota \nabla B ) \\ \nonumber
& = & \frac{1}{2} \left ( \partial_t \mathbf{E}^2 + \partial_t B^2 \right ) + B \nabla \cdot ( \iota \mathbf{E} )+ \nabla B \cdot (\iota \mathbf{E}) - \partial_t ( \iota B \mathbf{E} ) - \frac{1}{2} \left( \nabla B^2  + \nabla \mathbf{E}^2 \right ) \\ \nonumber
& &  - \left ( \mathbf{E} \wedge \nabla \right ) \mathbf{E} + 2 \left ( \nabla \cdot \mathbf{E} \right ) \mathbf{E} -  (\nabla \wedge \mathbf{E}) \mathbf{E} - \iota ( \nabla \cdot \mathbf{E} ) B + \iota ( \mathbf{E} \cdot \nabla ) B \\ \nonumber
& = &  \partial_t u + \nabla \cdot \mathbf{S} - \partial_t \mathbf{S} - \nabla u + ( \mathbf{E} \cdot \nabla + \nabla \cdot \mathbf{E} ) \mathbf{E} +  \left ( \mathbf{E} \cdot \nabla  - \nabla \cdot \mathbf{E} \right ) \iota B . \nonumber
\eea

\subsubsection*{Plane wave solution}

We take a trial solution
\be \label{TrialSolnLeadingC}
\psi(X) =  C \rme^{ \iota K \cdot X } = C \rme^{ \iota ( \mathbf{k} \cdot \mathbf{x} - w t ) } ,
\ee
where $ C $ is some constant multivector, then on substitution into Eq.~(\ref{Dirac3DTime}), we find
\bea \label{Dirac3DTrial}
\left ( \partial_t + c \mathbf{\nabla} \right ) C \rme^{ \iota ( \mathbf{k} \cdot \mathbf{x} - w t ) } & = &
 C ( - \iota w  ) \rme^{ \iGA K \cdot X } + c e_1 C \iota k_x \rme^{ \iGA K \cdot X }+ c e_2 C \iota k_y \rme^{ \iGA K \cdot X } \\ \nonumber
& = &  - w C \iota \rme^{ \iGA K \cdot X } + c \mathbf{k} C \iota \rme^{ \iGA K \cdot X } \\ \nonumber
& = &  - w C \iota \rme^{ \iGA K \cdot X } + c \mathbf{k} C \iota \rme^{ \iGA K \cdot X } \\ \nonumber
& = & \frac{- \iota m c^2}{\hbar}  C \rme^{ \iGA K \cdot X } \nonumber .
\eea
Multiplying from the right by $ \hbar \rme^{ -\iGA K \cdot X } $, we find  $ \left (\hbar w - c \hbar \mathbf{k} \right  )  C \iota = m c^2 \, \iota C $. 
We thus need to satisfy
\be \label{Dirac3DTrialSolve}
\left (  \hbar w  - c \hbar \mathbf{k} \right ) C+ m c^2  \iota C \iota =  0.
\ee
Given the multivector $ C = a + \mathbf{u} + \iota b $, we find $ \iota C \iota = - a + \mathbf{u} - \iota b $.  Hence we have the equation
\be \label{PlaneWaveEquationDirac}
\left (  \hbar w  - c \hbar \mathbf{k} \right )\left( a + \mathbf{u} + \iota b  \right ) = m c^2 \left (a - \mathbf{u} + \iota b \right ) . 
\ee
For a particle at rest we have $ E = \hbar w = m c^2 $, giving
\be
E \left( a + \mathbf{u} + \iota b  \right ) = m c^2 \left (a - \mathbf{u} + \iota b \right ).
\ee
Hence for positive energy solutions we require $ C = a + \iota b $, and for negative energy solutions with $ E = - \hbar w $ we require $ C = \mathbf{u} $.  Hence we have the positive and negative wavefunctions $ \psi^{+} = (a + \iota b) \rme^{ - \iGA w t } $ and $ \psi^{-} = \mathbf{u} \rme^{- \iGA w t } $.  The positive wavefunction acting on vector will rotate it clockwise and the negative wave function will rotate a vector in the negative direction in agreement with the de~Broglie formula $ E = \hbar w $.  Hence this model gives immediately the result that a negative energy, is a negative angular frequency which implies a negative time direction, with an inverted spin, in agreement with Feynman's interpretation of negative energies as particles traveling back in time with an inverted spin.

For the general case we need to equate scalar, vector and bivector components of Eq.~(\ref{PlaneWaveEquationDirac}) to find
\bea \label{DiracFourEquations}
-c \mathbf{u} \cdot \mathbf{p} + (\hbar w - m c^2 ) a & = & 0 \\ \nonumber
-a  c \mathbf{p} + (\hbar w + m c^2 ) \mathbf{u} + b c \iota \mathbf{p} & = & 0 \\ \nonumber
b (\hbar w - m c^2 ) \iota + c \mathbf{u} \wedge \mathbf{p} & = & 0 . \nonumber
\eea
From the first and third equations we find 
\be \label{solvingab}
 a = \frac{ c \mathbf{p} \cdot \mathbf{u} }{E - m c^2 } , \,\,\,\,  b = \frac{ c \mathbf{u} \wedge \mathbf{p} }{E - m c^2 } \iGA 
\ee
and substituting into the second equation gives
\be
-\frac{ c^2 \mathbf{p} \cdot \mathbf{u} }{E - m c^2 } \mathbf{p} + (E + m c^2 ) \mathbf{u} - \frac{ c^2 \mathbf{u} \wedge \mathbf{p} }{E - m c^2 }  \mathbf{p} = 0
\ee
multiplying through by $ E - m c^2  $ gives
\be
-c^2 \mathbf{p} \cdot \mathbf{u} \mathbf{p} + c^2 \mathbf{p}^2 \mathbf{u} - c^2 \mathbf{u} \wedge \mathbf{p} \mathbf{p} = 0
\ee
using the fact that $ E^2 - m^2 c^4 = c^2 \mathbf{p}^2 $, which can then be written
\be
-c^2 \mathbf{u} \mathbf{p} \mathbf{p} + c^2 \mathbf{p}^2 \mathbf{u} 
\ee
which is identically zero because using associativity $ (\mathbf{u} \mathbf{p}) \mathbf{p} = \mathbf{u} (\mathbf{p} \mathbf{p}) = \mathbf{u} \mathbf{p}^2 = \mathbf{p}^2 \mathbf{u} $.  Hence Eq.(\ref{solvingab}) is sufficient to satisfy solve Eq.~(\ref{DiracFourEquations}) and that allows both signs of energy $ E = \pm \hbar w $, and so we can write 
\be
C = a + \mathbf{u} + \iGA b = \frac{ c \mathbf{p} \cdot \mathbf{u} }{\pm E - m c^2 } + \mathbf{u} + \frac{ c \mathbf{p} \wedge \mathbf{u} }{\pm E - m c^2 } = \frac{ c \mathbf{p} \mathbf{u} }{\pm E - m c^2 } + \mathbf{u} = \left ( \frac{ c \mathbf{p} }{\pm E - m c^2 } + 1 \right ) \mathbf{u} ,
\ee
and substituting into Eq.~(\ref{TrialSolnLeadingC}) we have finally 
\be
\psi = \left ( \frac{ c \mathbf{p} }{\pm E - m c^2 } + 1 \right ) \mathbf{u} \rme^{ \iota ( \mathbf{k} \cdot \mathbf{x} - w t ) } .
\ee
Interpreted as an operator we can identify a rotation, reflection and a boost.

If we assume minimal coupling of the form  $ \mathbf{p} = - \iota \hbar \nabla - q \mathbf{A} $ and $ \mathbf{E} = \iota \hbar \partial_t - q V $ then we find
\be \label{DiracIn3DPsiCoupledAppendix}
\BoxOp \psi = \frac{m c}{\hbar} \psi  - \frac{q}{\hbar c} ( V + c \mathbf{A} ) \iota \psi \iota = \frac{m c}{\hbar} \psi  - \frac{q}{\hbar c} A \psi \iota,
\ee
where $ A = (V + c \mathbf{A} ) \iota  $ is the potential multivector, which can be compared with the momentum multivector in Eq.~(\ref{momentumMultivector}), corresponding to the four-potential if using this formalism.
This definition for the potential is compatible with Maxwell's equations, using
\be
F = - \BoxOp A = -\iota \left ( \frac{1}{c} \partial_t + \nabla \right ) \left ( V + c \mathbf{A} \right ) \iota = - \nabla V  - \frac{\partial \mathbf{A}}{\partial t} + c \nabla \wedge \mathbf{A} + \frac{ \partial V }{c \partial t } + c \nabla \cdot \mathbf{A} = \mathbf{E} + \iota c B ,
\ee
where $ \mathbf{E} = - \nabla V  - \frac{\partial \mathbf{A}}{\partial t} $ and $ \iota c B =  c \nabla \wedge \mathbf{A} $ and $ \frac{ \partial V }{c \partial t } + c \nabla \cdot \mathbf{A} =0 $ is the Lorenz gauge, which produces Maxwell's equations in terms of electromagnetic potentials as $ \BoxOp^2 \mathbf{A} = -J $, which are a set of three uncoupled Poisson equations.
We find $ F^2 = \mathbf{E}^2 - B^2 $ and $ J \cdot A = q V - q \mathbf{v} \cdot \mathbf{A} $, and so we have the field Lagrangian $ {\cal L }  = \frac{1}{2} F^2 + J \cdot A = \frac{1}{2} \left ( \mathbf{E}^2 - B^2 \right ) - \rho V + \mathbf{J} \cdot \mathbf{A} $, with Lagrange's equations $ \BoxOp \left ( \frac{ \partial {\cal L }}{\partial \left ( \BoxOp A \right ) } \right) - \frac{ \partial {\cal L }}{\partial A} = 0 $.

We find $ T = F \iota F = - \left (\mathbf{E}^2 + B^2 + 2 \iota B \mathbf{E} \right ) \iota = -\left ( U + \mathbf{S} \right ) \iota $, where $ U = \mathbf{E}^2 + B^2 $ is the total field energy and $ \mathbf{S} = 2 \iota B \mathbf{E} $ is the Poynting vector.  We can then write Poynting's energy conservation theorem as $ \BoxOp \cdot T = J \cdot F $.

\subsubsection*{Stationary state solutions}

We assume $ \mathbf{A} $ and $ V $ are time-independent and we look for stationary states solutions of the form
\be \label{StationaryStateTrialSolnLeadingC}
\psi(X) =  \psi(\mathbf{r}) \rme^{ -\iota w t }  =  \psi(\mathbf{r}) \rme^{ -\iota E t /\hbar} ,
\ee
and substituting into the Dirac equation in Eq.~(\ref{DiracIn3DPsiCoupledAppendix}) we find
\bea \label{StationaryStateAppendix}
\hbar c \BoxOp \psi & = & \iota \hbar \left ( \partial_t + c \nabla \right ) \psi(\mathbf{r}) \rme^{ -\iota w t } \\ \nonumber
& = & \iota \hbar \psi(\mathbf{r}) (-\iota w )\rme^{ -\iota w t }  + \iota \hbar c \nabla \psi(\mathbf{r}) \rme^{ -\iota w t } \\ \nonumber
& = & m c^2  \psi(\mathbf{r}) \rme^{ -\iota w t } - q( V + c \mathbf{A})\iota \psi(\mathbf{r}) \rme^{ -\iota w t } \iota . \nonumber
\eea
Multiplying from the right with $ \rme^{\iota w t } $, we find
\be
E \psi + \hbar c \nabla \psi \iota = -m c^2 \iota \psi \iota + q  ( V  - c \mathbf{A}) \psi  .
\ee
Writing $ \psi = a + \mathbf{v} + \iota b = \phi_A + e_1 \phi_2 = \phi_A + \phi_B $ which splits into the even and odd parts of the multivector, where $ \phi_A = a + \iota b $ and $ \phi_B = e_1 (c + \iota d )$, then we find two coupled equations
\bea \label{coupledDiracPDES}
E \phi_A + \hbar c \nabla \phi_B \iota & = &  (q  V + m c^2 ) \phi_A - q  c \mathbf{A} \phi_B   \\ \nonumber
E \phi_B + \hbar c \nabla \phi_A \iota & = & (q V - m c^2 ) \phi_B - q  c \mathbf{A} \phi_A \nonumber,
\eea
which correspond to conventional solutions \citep{bransden2000quantummechanics}.
From the second equation we find
\be
\phi_B = -\frac{q  c \mathbf{A} \phi_A + \hbar c \nabla \phi_A \iota}{E  -q V + m c^2 } = \frac{(-q  c \mathbf{A} + \hbar c \iota \nabla ) \phi_A }{E  -q V + m c^2 },
\ee
remembering that $ \iota $ commutes with $ \phi_A $ as it is the even subalgebra.

\subsubsection*{Non-relativisitic form of the Dirac equation}

For positive energy and non-relativistic speeds we have $ q V << m c^2 $ and $ p = \hbar k << m c^2 $ and so we have approximately
\be \label{phiBSimplified}
\phi_B = \frac{(-q  \mathbf{A} + \hbar \iota \nabla ) \phi_A }{2 m c },
\ee
so that $ \phi_B \approx \frac{\hbar \mathbf{k}}{m c} \phi_A \approx \frac{\mathbf{v}}{c} \phi_A $ and so for non-relativistic speeds $ \phi_B << \phi_A $.  Letting $ E = E' + m c^2 $, where $ E' << m c^2 $, and substituting Eq.~(\ref{phiBSimplified}) back into Eq.~(\ref{coupledDiracPDES}) we find
\be
(E' - q V) \phi_A =   c \left ( q  \mathbf{A} + \hbar \iota \nabla  \right) \frac{(q  \mathbf{A} - \hbar \iota \nabla) }{2 m c}  \phi_A ,
\ee
which expands to
\bea
2m (E' - q V) \phi_A & = &  q^2  \mathbf{A}^2 \phi_A - \hbar^2  \nabla^2 \phi_A - q  \hbar  \mathbf{A} \iota \nabla \phi_A + q \hbar \iota \nabla \left ( \mathbf{A} \phi_A  \right )    \\ \nonumber
& = & - \hbar^2  \nabla^2 \phi_A + q^2  \mathbf{A}^2 \phi_A + q \hbar \iota (\mathbf{A} \cdot \nabla ) \phi_A + q \hbar \iota (\mathbf{A} \wedge  \nabla ) \phi_A  \\ \nonumber
& & + q \hbar \iota (\nabla  \cdot \mathbf{A}) \phi_A + q \hbar \iota  (\nabla \wedge \mathbf{A}) \phi_A  + q \hbar \iota \dot{\nabla} \mathbf{A} \dot{\phi_A}    \\ \nonumber
& = & \left (- \hbar^2  \nabla^2 \phi_A + q^2  \mathbf{A}^2 + q \hbar \iota \mathbf{A} \cdot \nabla  + q \hbar \iota \nabla  \cdot \mathbf{A} \right ) \phi_A  \\ \nonumber
& & + q \hbar \iota \left ( \mathbf{A} \wedge  \nabla +  \nabla \wedge \mathbf{A} \right ) \phi_A  + q \hbar \iota \dot{\nabla} \mathbf{A} \dot{\phi_A}  ,  \nonumber
\eea
using associativity $ A \left ( \nabla \phi_A  \right ) = \left ( A \nabla \right ) \phi_A   $ and  $ \nabla \left ( \mathbf{A} \phi_A  \right ) = \nabla \mathbf{A} \phi_A  + \dot{\nabla} \mathbf{A} \dot{\phi_A} $. 
However $ \dot{\nabla} \mathbf{A} \dot{\phi_A} = ( \dot{\nabla} \cdot \mathbf{A} + \dot{\nabla} \wedge \mathbf{A} ) \dot{\phi_A} = \mathbf{A} \cdot \nabla \phi_A - \mathbf{A} \wedge \nabla \phi_A $, which gives a cancellation to
\bea
2m (E' - q V) \phi_A  & = & \left (- \hbar^2  \nabla^2 \phi_A + q^2  \mathbf{A}^2 + q \hbar \iota \mathbf{A} \cdot \nabla  + q \hbar \iota \nabla  \cdot \mathbf{A} \right ) \phi_A  \\ \nonumber
& & + q \hbar \iota \left ( \nabla \wedge \mathbf{A} \right ) \phi_A + q \hbar \iota \left ( \mathbf{A} \cdot \nabla \right ) \phi_A . \nonumber
\eea
Now in two dimensions $ \iota B_z = \nabla \wedge A $ 
\bea
2m (E' - q V) \phi_A  & = & \left (- \hbar^2  \nabla^2 \phi_A + q^2  \mathbf{A}^2 + q \hbar \iota \mathbf{A} \cdot \nabla  + q \hbar \iota \nabla  \cdot \mathbf{A} \right ) \phi_A  \\ \nonumber
& & - q \hbar B_z \phi_A + q \hbar \iota \left ( \mathbf{A} \cdot \nabla \right ) \phi_A  . \nonumber
\eea
The term in brackets in three dimensions factorizes to $ ( i \hbar \nabla + q \mathbf{A})^2 = - \hbar^2 \nabla^2 + q \mathbf{A}^2 + q \hbar i \nabla \cdot \mathbf{A} + q \hbar i \mathbf{A} \cdot \nabla  $, however in two dimensions the pseudoscalar is non-commuting and so this factorization is not possible.  We can select the Coulomb gauge, $ \nabla \cdot \mathbf{A} = 0 $ and for an electron $ q = -e $ to give
\be \label{PauliGA2D}
E' \phi_A   =  \left ( \frac{1}{2 m} (- \hbar^2  \nabla^2 + e^2  \mathbf{A}^2 - 2 e \hbar \iota \mathbf{A} \cdot \nabla  +e \hbar B_z) -e V \right ) \phi_A ,
\ee
which is identical to the Pauli equation in two-dimensions, typically written as
\be
E' \psi_A   =   \left ( \frac{1}{2 m}(- \hbar^2  \nabla^2 + e^2  \mathbf{A}^2 - e \hbar \iota \mathbf{A} \cdot \nabla +e \hbar \sigma \cdot \mathbf{B}) -e V \right ) \psi_A ,
\ee
except for the extra factor of two on the $ \mathbf{A} \cdot \nabla $ term, where $ \psi_A $ is the conventional Pauli spinor and $ \mathbf{\sigma} $ is the three-vector of Pauli matrices.  This could be an artifact of a two-dimensional form.  As is well known the coefficient $ \frac{e \hbar}{2 m } $ in front of $ B_z $, gives a spin gyromagnetic ratio $ g_s = 2 $ in close agreement with experiment.

If we select $ \mathbf{A} = 0 $ then Eq.~(\ref{PauliGA2D}) reduces to
\be
 \left (- \frac{\hbar^2}{2 m} \nabla^2 -e V \right ) \psi_A = E' \psi_A,
\ee
the Schr{\"o}dinger equation in two dimensions.

The representation of basic physical equations in a two-dimensional Clifford algebra $ {\rm{Cl}}_{2,0} $, representing a uniform positive signature, gains significance from the isomorphism
\be
{\rm{Cl}}_{(n+2,0)} \approx {\rm{Cl}}_{(0,n)} \otimes {\rm{Cl}}_{(2,0)}
\ee
so that higher dimensional Clifford algebra can be constructed from the two-dimensional case.  This result also leads to the well known Bott periodicity of period eight for real Clifford algebras \citep{gallier2005clifford}.

\subsection*{The multivector model for the electron}

\begin{figure}[htb]

\begin{center}
\includegraphics[width=3.1in]{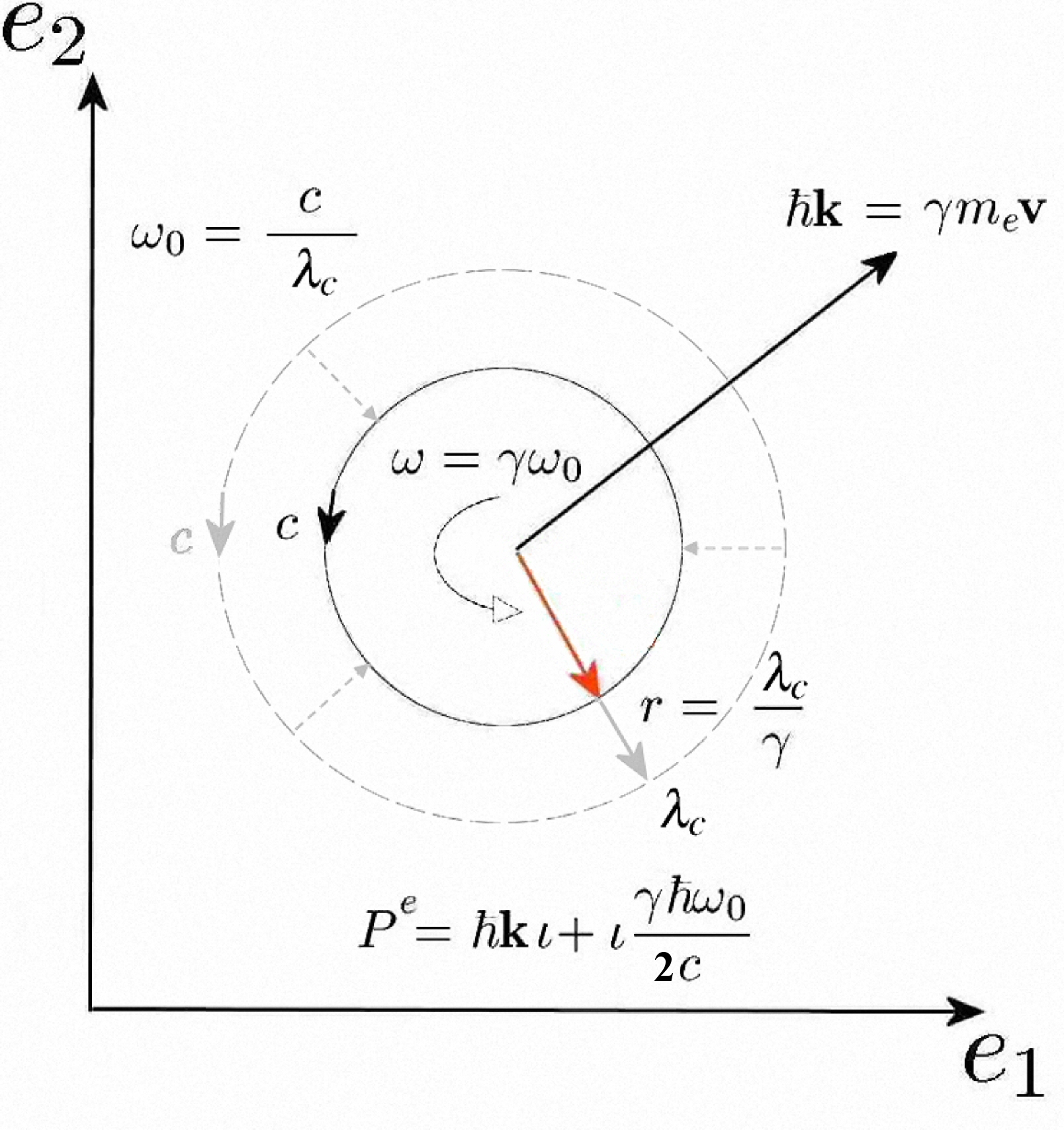}
\end{center}

\caption{Multivector model for the electron, consisting of a light-like particle orbiting at the de~Broglie angular frequency $ \omega_0 $ at a radius of $ r_0 = \lambda_c/2 $ in the rest frame, and when in motion described generally by the multivector  $ P_e = \hbar \mathbf{k} \iota + \iota \frac{\gamma \hbar \omega_0 }{ 2 c} $.  Under a boost, the de~Broglie angular frequency will increase to $ \gamma \omega_0 $, giving an apparent mass increase and time dilation, the electron radius will also shrink by $ \gamma $, implying length contraction, thus naturally producing the key results of special relativity. \label{ElectronBivector}}
\end{figure}

We know from inspection of Eq.~(\ref{electronMultivector}), that under a boost, the frequency $ \omega_0 $ will increase to $ \omega = \gamma \omega_0 $, with the radius required to shrink to $ r = r_0/\gamma $, so that the tangential velocity $ v = r w = \left( \frac{r_0}{\gamma} \right ) \gamma w_0 = r_0 w_0 =  c $, remains at the speed of light.  
Hence this simplified two-dimensional model in Fig. \ref{ElectronBivector}, indicates that under a boost,  the de~Broglie frequency will increase to $ \gamma \omega_0 $ implying an energy and hence a mass increase $ \gamma m_0 $, the frequency increase also implies time dilation, and the shrinking radius producing length contraction, thus producing known relativistic effects.

Now because the wave vector term $ \mathbf{k} \iota $ in Eq.~(\ref{electronMultivector}) represents a momentum perpendicular to the direction of motion, we can identify it with the angular momentum of the lightlike particle with $ p = E/c = \gamma \hbar w_0/(2 c) = \gamma m c $, remembering $ \mathbf{r} \cdot \mathbf{p} = 0  $, we have the spin angular momentum $ L = \mathbf{r} \mathbf{p} = \mathbf{r} \wedge \mathbf{p} = [ \mathbf{r} , \mathbf{p} ] = \iota \left ( \frac{\hbar}{2 \gamma m c } \right ) \left ( \gamma m c \right ) = \iota \frac{\hbar}{2} $ which is invariant, as expected for a spin-$\frac{1}{2}$ particle.

Integrating the momentum multivector with respect to the proper time $ \tau $, remembering that $ d t = \gamma d \tau $, and dividing by the rest mass $ m $ we find
\be \label{electronRotation}
X =  \mathbf{x} \iota + \iota \frac{\hbar \theta_0}{2 m c} =  \mathbf{x} \iota  + \iota r_0 \theta_0,
\ee
where $ \omega_0 = \frac{ d \theta_0}{d t} $. Inspecting the bivector component, we find
\be
r_0 d \theta_0  = r_0 w_0 d t =  \left ( \frac{\hbar}{2 m c}\right ) \left ( \frac{2 m c^2 }{\hbar} \right ) d t = c d t ,
\ee
and hence the time can be identified as the circumferential distance $ r_0 d \theta_0 $ at half the Compton radius $ \frac{\hbar}{2 m c} $.  

\subsection*{Wave mechanics}

We now continue to work within the two-dimensional Clifford multivector to intuitively produce two-dimensional versions of Dirac's and Maxwell's equations.
From the de~Broglie hypothesis that all matter has an associated wave \citep{broglie1923thesis}, given by the relations $ \mathbf{p} = \hbar \mathbf{k} $ and $ E = \hbar w $, we find from Eq.~(\ref{momentumMultivector}) the wave multivector
\be \label{waveMultivector}
K = \frac{P}{\hbar} = \mathbf{k} \iota +  \frac{w}{c} \iota = \left ( \frac{w}{c}  + \mathbf{k} \right )\iota.
\ee
We now find the dot product of the wave and spacetime multivectors $ K \cdot X = \mathbf{k} \cdot \mathbf{x} - w t  $, giving the phase of a traveling wave. 
Hence for a plane monochromatic wave we can write $ \psi = \rme^{\iota K \cdot X } = \rme^{\iota (\mathbf{k} \cdot \mathbf{x} - w t) } $, which leads to the standard substitutions, $ \mathbf{p} = - \iota \hbar \nabla $ and $ \mathbf{E} = \iota \hbar \partial_t $ and so we define from the momentum multivector 
\be \label{BoxGradientMom}
\BoxOp = -\frac{\iota}{\hbar} P =  \frac{-\iota}{\hbar} \left ( \mathbf{p} \iota +  \frac{E}{c} \iGA \right ) = \iota \left ( \frac{1}{c} \partial_t + \nabla   \right ) ,
\ee
where $ \nabla = e_1 \partial_x + e_2 \partial_y  $.  We therefore find $ -\BoxOp^2 = \frac{1}{c^2} \partial_t^2 - \nabla^2 $ the d'Alembertian in two dimensions, so that $ \BoxOp $ is the square root of the d'Alembertian. Following Dirac, we therefore write
\be \label{Dirac3DTime}
\BoxOp \psi  = \frac{m c }{\hbar} \psi ,
\ee
which is isomorphic to the conventional Dirac equation (see Appendix), and comparable to the Dirac equation previously developed in three dimensional Clifford algebra \citep{hestenes2003mysteries,Boudet}.  Acting from the left a second time with the differential operator $ \BoxOp  $ we produce the Klein-Gordon equation, $ \left (\frac{1}{c^2} \partial_t^2 - \nabla^2  \right ) \psi = -\frac{m^2 c^2}{\hbar^2 } \psi $. 

For the non-relativistic case, summing the kinetic and potential energy, we find the total energy $ E = T + V = \frac{p^2}{2 m } + V $, and substituting the standard operators for $ \mathbf{p} $ and $ E $ we find 
\be \label{Schrodinger}
\left (-\frac{\hbar^2}{2 m } \nabla^2 + V \right ) \psi = \iota \hbar \partial_t \psi ,
\ee
which for a multivector $ \psi = a + \iota b $ produces the Schr{\"o}dinger equation in two dimensions, remembering that $ \iota $ commutes with scalars and so acts equivalently to the scalar imaginary $ \sqrt{-1} $ in this case. The wave function represents a rotation in the plane $ e_1 e_2 $ and therefore the Schr{\"o}dinger equation as represented in Eq.~(\ref{Schrodinger}) describes an eigenstate of spin in the plane $ e_1 e_2 $.

\subsubsection*{Maxwell's equations}

For a massless particle we have the Klein-Gordon equation $ \BoxOp^2 \psi = 0 $, with a solution $ \BoxOp \psi = 0 $ from the Dirac equation in Eq.~(\ref{Dirac3DTime}), however we can also write
\be \label{MaxwellPrimitive}
\BoxOp \psi = J,
\ee
where $ J $ is a general multivector. Acting a second time with the spacetime gradient produces
\be
\BoxOp^2 \psi = \BoxOp J
\ee
and provided $ \BoxOp J = 0 $, we satisfy the massless Klein-Gordon equation.
Now $ \BoxOp J = \BoxOp \cdot J + \BoxOp \wedge J $, and for $ J = ( \rho + \mathbf{J} ) \iota  $ representing source currents, where we now switch to natural units with $ c = \hbar = 1 $, we find firstly
\be
\BoxOp \cdot J = \partial_t \rho + \nabla \cdot \mathbf{J}  = 0
\ee
which is the requirement of charge conservation. Also $ \BoxOp \wedge \mathbf{J} = \partial_t \mathbf{J} + \nabla \wedge \mathbf{J} + \nabla \rho $ that is also zero for steady currents and curl free sources, and hence for this restricted case Eq.~(\ref{MaxwellPrimitive}) is also a solution to the Klein-Gordon equation. Then writing the electromagnetic field as the multivector $ \psi = \mathbf{E} + \iota B $, we have produced Maxwell's equations, that is, from Eq.~(\ref{MaxwellPrimitive}) we find
\be
\left ( \partial_t + \nabla \right ) \left (\mathbf{E} + \iota B \right ) =   \rho - \mathbf{J}   
\ee
that when expanded into scalar, vector and bivector components, gives $ \mathbf{\nabla} \cdot \mathbf{E} = \rho $, $ \iota \mathbf{\nabla} B - \frac{\partial \mathbf{E}}{ \partial t}  =  \mathbf{J} $, and $ -\iota \mathbf{\nabla} \wedge \mathbf{E} + \frac{\partial B}{ \partial t}  = 0  $ respectively, and noting that in three dimensions $ -\iota \mathbf{\nabla} \wedge \mathbf{E} = \nabla \times \mathbf{E} $ and $ \iota \nabla B = \nabla \times \mathbf{B} $, we see that we have produced Maxwell's equations for the plane. This equation also very naturally expands to produce Maxwell's equations in three dimensions as  $ \left ( \partial_t + \nabla \right ) \left (\mathbf{E} + i \mathbf{B}  \right ) =   \rho - \mathbf{J} $, where the magnetic field now becomes a three-vector, with $ i = e_1 e_2 e_3 $ the trivector \citep{Boudet}.

Hence using a general multivector $ \psi = \lambda + \mathbf{E} + \iota B $ we have in natural units
\bea \label{MaxwellDiracKleinGordon}
\BoxOp^2 \psi & = & m^2 \psi , \,\,\,\,\,\,\,\, \rm{Klein-Gordon} \,\, \rm{equation} \\ \nonumber
\BoxOp \psi & = & m \psi , \,\,\,\,\,\,\,\, \rm{Dirac} \,\, \rm{equation} \\ \nonumber
\BoxOp \psi & = &  J , \,\,\,\,\,\,\,\,\,\, {\rm{Maxwell's}} \,\, {\rm{equations}} \,\, (m=0) . \nonumber
\eea
That is, with the assumption of the form of the spacetime gradient $ \BoxOp $ given in Eq.~(\ref{BoxGradient} and the spacetime multivector given by Eq.~(\ref{spacetimeEvent}), the two simplest first order differential equations we can write are Maxwell's equations and the Dirac equation. Also, setting $ m = 0 $ in the Dirac equation we find  $ \BoxOp \psi = 0 $, the Weyl equation for the plane.   Hence the two dimensional Clifford multivector provides a natural `sandbox', or simplified setting, within which to explore the laws of physics.  As a generalization, Maxwell's source term can be expanded to a full multivector to give $ J = ( \rho + \mathbf{J} + \iota s ) \iota $, and $ s $ describes magnetic monopole sources.

We now find the form of the homogeneous Lorentz group for these multivectors.

\subsubsection*{Solutions}

An elegant solution path is found for the Maxwell and Dirac equation in Eq.~(\ref{MaxwellDiracKleinGordon}) through defining the field $ \psi $ in terms of a multivector potential $ A = \iota \left ( -V + c \mathbf{A} + \iota M \right ) $, with $ M $ describing a possible monopole potential, given by
\be \label{PotentialDefn}
\psi = \BoxOp A .
\ee
We then find Maxwell's equations defined in Eq.~(\ref{MaxwellDiracKleinGordon}) in terms of a potential becomes
\be
\BoxOp^2 A = J .
\ee
Now, as $ \BoxOp^2 = \nabla^2  - \frac{1}{c^2} \partial_t^2  $ is a scalar differential operator we have succeeded in separating Maxwell's equations into four independent inhomogeneous wave equations, given by the scalar, vector and bivector components of the multivectors, each of which have the well known solution \cite{Griffiths:1999} given by
\be
A = \frac{\mu_0}{4 \pi} \int_{vol} \frac{J}{r} d \tau ,
\ee
where $ r = | \mathbf{r} - \mathbf{s}| $, the distance from the field point $ \mathbf{r} $ to the charge at $ \mathbf{s}
 $, and where we calculate values at the retarded time.
The field can then be found from Eq.~(\ref{PotentialDefn}) by differentiation.

We find $ \iota ( \partial_t + \nabla ) \left ( u + \mathbf{S} \right ) \iota = - \partial_t u  - \nabla \cdot \mathbf{S} + \partial_t \mathbf{S} + \nabla u - \nabla \wedge \mathbf{S} $, therefore we can express the conservation of momentum and energy from Eq.~(\ref{FJEqn}) and Eq.~(\ref{FJFieldsEqn}) as
\be
\partial T = -\iota \psi J + ( \mathbf{E} \cdot \nabla + \nabla \cdot \mathbf{E} ) \mathbf{E} + 2 \iota \nabla B \cdot \mathbf{E} ,
\ee
where the scalar components express the conservation of energy and the vector components the conservation of momentum.  The cumbersome $ ( \mathbf{E} \cdot \nabla + \nabla \cdot \mathbf{E} ) \mathbf{E} $ term is typically absorbed into a stress energy tensor \citep{Griffiths:1999}. The conservation of charge $ \partial_t \rho + \nabla \cdot \mathbf{J} = 0 $ also follows from Maxwell's equation through taking the divergence of Eq.~(\ref{MaxwellPrimitiveOne}).

We can also define a Lagrangian $ L = \frac{1}{2} \epsilon_0 \psi^2 - J \cdot A $, which through the Euler-Lagrange equations produces Eq.~(\ref{MaxwellPrimitiveOne}).

%\section*{Acknowledgments}

%\section*{References}
% The bibtex filename
\bibliography{quantum}

\end{document}